\documentclass[10pt,a4paper]{article}%
\usepackage{natbib}
\usepackage{a4wide}

\usepackage[sc]{mathpazo}
\usepackage[T1]{fontenc}

\usepackage{amsmath}
\usepackage{amsfonts}
\usepackage{amssymb}
\usepackage{amsthm}
\usepackage{enumerate}

\usepackage{microtype}

\usepackage{graphicx}
	\setkeys{Gin}{width=1\textwidth}

\usepackage{ifpdf}
\ifpdf
\usepackage{epstopdf}
\else
\fi

\usepackage[english]{babel}
\usepackage[latin1]{inputenc}
\usepackage{mathrsfs}

\usepackage[x11names]{xcolor}

\usepackage{hyperref}
	\hypersetup{pdfauthor=Tristan LAUNAY,
				pdfstartview=FitH, pdfstartpage=1, bookmarksopen=true,
		 bookmarksopenlevel=0, bookmarksnumbered=true, naturalnames=true, breaklinks=false, linktocpage=true}
\usepackage[all]{hypcap}

\usepackage{dsfont}

\usepackage{xspace}

\DeclareMathOperator*{\var}{var}

\newcommand{\R}{\mathbb{R}}

\newcommand{\cA}{\mathcal{H}}
\newcommand{\cB}{\mathcal{}}

\newcommand{\one}{\mathds{1}}

\newcommand{\esp}{\mathbb{E}}

\newcommand{\loi}[1]{\mathcal{#1}}

\newcommand{\ud}{\,\mathrm{d}}

\newcommand{\interff}[2]{[#1,\,#2]}
\newcommand{\interoo}[2]{]#1,\,#2[}

\newcommand{\interfo}[2]{[#1,\,#2[}

\newtheorem{theorem}{Theorem}

\newtheorem{example}[theorem]{Example}

\newtheorem{property}[theorem]{Property}
\newtheorem{proposition}[theorem]{Proposition}
\newtheorem{remark}[theorem]{Remark}

\begin{document}

\title{Construction of an informative hierarchical prior for a small sample with the help of historical data and application to electricity load forecasting} 
\author{Tristan Launay\(^{1,2}\) \and Anne Philippe\(^1\) \and Sophie Lamarche\(^2\)}
\maketitle
\footnotetext[1]{Laboratoire de Math\'ematiques Jean Leray, 2 Rue de la Houssini\`ere -- BP 92208, 44322 Nantes Cedex 3, France}
\footnotetext[2]{Electricit\'e de France R\&D, 1 Avenue du G\'en\'eral de Gaulle, 92141 Clamart Cedex, France}
\begin{abstract}
We are interested in the estimation and prediction of a parametric model on a short dataset upon which it is expected to overfit and perform badly. To overcome the lack of data (relatively to the dimension of the model) we propose the construction of an informative hierarchical Bayesian prior based upon another longer dataset which is assumed to share some similarities with the original, short dataset. We illustrate the performance of our prior on simulated dataset from three standard models. 
Then we apply the methodology to a working model for the electricity load forecasting on real datasets, where it leads to a substantial improvement of the quality of the predictions.
\medskip

\end{abstract}

\section{Introduction}\label{sec:intro}
We consider the problem of building a prior distribution when only a small number of observations is available. In this context the choice of the prior becomes crucial, having a strong effect on the inference. However it can also be a good tool to bring information so as to improve the estimations in complicated problems (e.g. model in high dimension with respect to the number of observations). Using the ideas of hierarchical modelling, we propose a parametric prior model, that requires the knowledge of another long historical sample \(y^\cA\) upon which the model performs equally well in both estimation and prediction. We will assume that this historical sample shares some similarities with our short sample of interest \(y^\cB\). 

The study is motivated by the modelling and forecasting of electricity load which is a problem well-known within both the academic and the applied statistics community \cite[see e.g.][]{BunnFarmer}. We are interested in the development of a methodology to improve the estimation and the predictions of a parametric multi-equation model (similar to the one presented in \cite{Bruhns}) over a short dataset. The limited size of the dataset coupled with the high dimensionality of the model leads to a typical overfitting situation when using the maximum likelihood approach: the fitted values are relatively close to the observations while the errors in prediction are an order of magnitude larger or more (due to the very periodic nature of the regressors involved, the model typically requires 4 or 5 years of data to provide satisfactory predictions). This overfitting behaviour can be somewhat alleviated by the use of a Bayesian estimation relying on an informative prior distribution, but the very fact that the data available is limited makes the posterior distribution all the more sensitive to the choice of that prior. Although electricity load curves may largely differ from one population to another, they may also share some common features. The latter case is expected to happen when the global population studied is an aggregation of non-homogeneous subpopulations for which the estimations are made harder due to the relative lack of data.

Different strategies can be found in the literature of hierarchical model to choose or to estimate the parameters of the prior distribution. See \cite{Gelmanetal,GelmanHill,congdon2010} for a general review on the subject of hierarchical models. These parameters can be assumed unknown and a new level is added to the hierarchical structure. This commonly used approach increases the dimension of the model, which is generally not recommended when dealing with small samples.

In \cite{Gelmanetal}, Chapter 5, some methods using historical samples are presented and illustrated on several examples. Typically, both samples are supposed to be coming from a common model. Assuming that the prior belongs in a parametric family, the hyperparameters of the prior are chosen as the estimated values on the historical sample. This strategy is not robust to small evolutions of the parameters between the historical and the short samples, leading to a bias in the estimation as illustrated in Section \ref{sec:electricity}. Note that it is also possible to estimate the prior on the sample itself, the approaches are often called empirical Bayesian method. This is not a full Bayesian analysis but the approach can be viewed as an approximation to hierarchical models (see e.g. \cite{Efron-empirical-Bayes} for details). 

Instead of estimating the parameter from historical sample, one can define a joint model on the samples $(y^\cA,y^\cB)$ where both samples come from the same parametric model but have different parameters (denoted respectively $\theta_\cA$ and $\theta_\cB$). The prior distribution on $(\theta_\cA,\theta_\cB)$ conveys the link and the closeness between both parameters with the help of a dependence structure. \cite{bouveyron2010adaptive} propose this method to estimate regression mixtures when the population has changed between the training and the predictive stages. They choose a conditional distribution of $\theta_\cB$ given $\theta_\cA$ which is centered in $\theta_\cA$. This approach requires knowledge of the joint distribution of samples $(y^\cA,y^\cB)$, which are assumed independent in \cite{bouveyron2010adaptive}. This leads to a transfer learning in both directions and attributes a higher importance to the largest historical sample. 

In this paper, we assume that the long and the short datasets are somehow similar in a non-obvious way. That the similarity between the parameters underlying the two datasets (we will assume they are indeed coming from the model we consider) cannot be easily guessed prevents us from trying to model the datasets simultaneously because it would require a rather precise knowledge of the link between the two. We propose a general way of building an informative hierarchical prior for the short dataset from the long historical one that goes as follows:
\begin{enumerate}
 \item we first estimate the posterior distribution on the long dataset using a non-informative prior, arguing that the design of an informative prior for this dataset is not necessary, since the data available are enough to estimate and predict the model in this case ;
 \item we extract key pieces of information from this estimation to design an informative prior for the short dataset which takes into account the prior information that the datasets are somehow similar, via the introduction of hyperparameters designed to model and estimate this similarity.
\end{enumerate}

The paper is organised as follows. In Section \ref{sec:methodo} we focus on the general methodology and propose a informative Gaussian prior model. From the two first moments, we show the transfer of information between the historical sample and the prior model. In Section \ref{sec:toyexamples} we illustrate the performance of our methodology on simulated dataset from three standard models : the auto-regression model, the polynomial regression and a Poisson model with random effects. In Section \ref{sec:electricity}, we introduce the model used for our main application of electricity load forecasting. We discuss the results obtained on French electricity datasets and show that our new methodology is competitive compared to three alternative standard methods. We also study the effect of the small dataset's sample size upon the predictive quality of the model and show that the informative prior provides reasonable forecasts even when the lack of data is severe. The ad hoc MCMC algorithms we developed to approximate numerically the various estimators are push backed into the Appendix (see Section \ref{sec:appendix}) so as not to obfuscate the main point of the paper by technical details. 

\section{Methodology}\label{sec:methodo}
\subsection{General principle}\label{subsec:generalprinciple}
We denote \(y^\cB\) the short sample over which we would like to estimate a parametric model $\{ f_\theta, \; \theta \in \Theta\}$ and to predict future observations. We denote \(y^\cA\) a historical sample coming from the same parametric model. We assume that this is a long dataset known to share some common features with \(y^\cB\). Let \(\theta_{\cA}\) and \(\theta_{\cB}\) be the corresponding parameters.
The parameter of interest is \(\theta_{\cB}\) . 

Let \(\pi_{\cA}\) and \(\pi_{\cB}\) be the prior distributions used to estimate \(\theta_{\cA}\) and \(\theta_{\cB}\) respectively. Let \(\pi_{\cA}(\cdot|y^{\cA})\) and \(\pi_{\cB}(\cdot|y^{\cB})\) be the associated posterior distributions. We will denote \(\widehat{\theta}_{\cA}\) and \(\widehat{\theta}_{\cB}\) the corresponding Bayes estimates (i.e. the posterior means).  

We propose a method to the improve parameter the estimations and model predictions for the dataset \(y^\cB\) with the help of the dataset \(y^\cA\). Observe first that the choice of \(\pi_{\cA}\) is not crucial as long as it remains non-informative enough because, \(\theta_{\cA}\) can be correctly estimated from the data alone on \(y^\cA\). The naive pick \(\pi_{\cB} = \pi_{\cA}(\cdot|y^{\cA})\) is not a viable solution as soon as the parameters of \(\theta_{\cA}\) and \(\theta_{\cB}\) differ since the variance of the posterior distribution \(\pi_{\cA}(\cdot|y^{\cA})\) is too small: in that case the few observations of \(y^\cB\) will not be able to make up for the difference between the posterior mean \(\widehat{\theta}_{\cA}\) and the true value of the parameter \(\theta_{\cB}\). Assuming that the parameters \(\theta_{\cA}\) and \(\theta_{\cB}\) are identical is too restrictive in practise. To allow for more flexibility we introduce hyperparameters to account for the similarity between the datasets. We now describe the informative hierarchical prior we designed.

Assume that the prior distribution \(\pi_{\cB}\) is to be chosen within the parametric family
\[\mathcal{F} = \{\pi_{\lambda} ;\; \lambda \in \Lambda \}.\]
Since selecting \(\pi_{\cB}\in\mathcal{F}\) is equivalent to picking \(\lambda^{\cB}\in\Lambda\), and since we would like \(\pi_{\cB}\) to retain some key-features of \(\pi_{\cA}(\cdot|y^{\cA})\), we want to pick \(\lambda^{\cB}\) using some of the information contained inside the posterior distribution \(\pi_{\cA}(\cdot|y^{\cA})\). We assume that there exists an operator \(T:\mathcal{F}\rightarrow\Lambda\), such that
\[T[\pi_{\lambda}] = \lambda,\]
and choose \(\lambda^{\cB}\) proportional to \(T[\pi_{\cA}(\cdot|y^{\cA})]\), in the sense that
\[\lambda^{\cB} = K T[\pi_{\cA}(\cdot|y^{\cA})],\] where \(K:\Lambda\rightarrow\Lambda\) itself is an unknown linear operator that we assume diagonal for ease of use.

The operator \(K\) can be interpreted as a similarity operator between \(\theta_{\cA}\) and \(\theta_{\cB}\), and its diagonal components as similarity coefficients measuring how close the two datasets \(y^\cA\) and \(y^\cB\) really are when looked at through the model. The diagonal components of \(K\) are hyperparameters of the prior we designed, and we give them a vague hierarchical prior distribution centred around \(q\), the prior on \(q\) being vague and centred around \(1\).

The hyperparameter \(q\) may also be regarded as a more global similarity coefficient, since it represents the mean of all the similarity coefficients. The prior mean of \(q\) is forced to \(1\) to reflect the prior knowledge that the datasets are somehow similar. The variance of the prior distribution of \(q\) could in theory be reduced, going from a vague prior to a more informative structure, depending on the confidence we have over the similarity between the datasets. We chose not to however, so as to keep the procedure we describe from requiring any delicate subjective adjustments.

\begin{remark}
 As we suppose that $y^\cA$ is a large sample, the posterior distribution \(\pi_{\cA}(\cdot|y^{\cA})\) can be approximated by Gaussian distribution under suitable regularity conditions.
Therefore we can replace \[\lambda^{\cB} = K T[\pi_{\cA}(\cdot|y^{\cA})],\] by \[\lambda^{\cB} = K T[ \mathcal{N} (\hat\theta_{\cA,n}^{ML}, I_{\cA,n}^{-1} )],\] where $ \hat\theta_{\cA,n}^{ML} $ is the maximum likelihood estimate and where $I_{\cA,n}$ is an estimate of the expected Fisher Information matrix calculated on the sample $y^\cA$ \cite[chap 4]{Ghosh}.
\end{remark}

We present now two frequent situations where the above procedure can be written in a simpler way.

\begin{example}[Method of Moments]
We assume that the elements of \(\mathcal{F}\) can be identified via their \(m\) first moments: the operator \(T\) can then be reduced to a function \(F\) of the \(m\) first moments operators, i.e. \(\lambda = T[\pi_{\lambda}] = F(\esp(\theta), \ldots, \esp(\theta^m))\). The expression of \(\lambda^{\cB}\) then becomes
\[\lambda^{\cB} = K F(\esp(\theta|y^{\cA}), \ldots, \esp(\theta^m|y^{\cA})).\]
Note that, if the prior requires the specification of at least the two first moments, even though the priors from the upper layers of the model are vague, the correlation matrix estimated on the dataset \(y^\cA\) remains untouched and is directly plugged into in the informative prior if we consider centred moments for orders greater than \(1\).
\end{example}

\begin{example}[Conjugacy]
We consider the case where \(\mathcal{F}\) is the family of priors conjugated for the model. If the prior \(\pi_{\cA}\) belongs to \(\mathcal{F}\) then the associated posterior distribution \(\pi_{\cA}(\cdot|y^{\cA})\) does too and we denote \(\lambda^{\cA}(y^{\cA})\) the parameter that corresponds to it. The expression of \(\lambda^{\cB}\) then reduces to
\[\lambda^{\cB} = K \lambda^{\cA}(y^{\cA}).\]
\end{example}

\subsection{Gaussian strategy}\label{Pgauss}
Under suitable conditions of regularity (see \cite{Ghosh}, \cite{Launay2}), the posterior distribution tends to look like a normal distribution asymptotically. This suggests the use of a normal distribution as the prior distribution $\pi_\cB$. Let us denote \(\widehat{\mu}^{\cA}\) and \(\widehat{\Sigma}^{\cA}\) the posterior mean and posterior variance of the parameter \(\theta_{\cA}\).

As described in the general methodology, we introduce new parameters to model the similarity between the two datasets, \((k, l) \in \R^d \times \R\) and \((q,r)\in \R\times\R_+^*\). The informative hierarchical prior is of the form 
\begin{align}
\theta_\cB | k,l &\sim \loi{N}(K\widehat{\mu}^{\cA}, l^{-1}\widehat{\Sigma}^{\cA}) \label{prioreta}\\
k|q,r &\sim \loi{N}(q(1,\ldots,1)^\prime, r^{-1} I_d) \label{priork}
\end{align}
where \(K=\text{diag}(k)\). Again, the coordinates of the vector \(k\) can be interpreted as similarity coefficients between \(\theta_{\cA}\) and \(\theta_{\cB}\) and the strictly positive scalar \(l\) can be seen as a way to alternatively weaken or strengthen the covariance matrix as needed. Hyperparameters \(q\) and \(r\) are more general indicators of how close \(\theta_{\cA}\) and \(\theta_{\cB}\) are, \(q\) corresponding to the mean of the coordinates of \(k\) and \(r\) being their inverse-variance. For a complete definition of the prior distribution, we of course need to specify the prior on \(l\), \(q\), \(r\). Based on conjugacy considerations, we use:
\begin{align}
l &\sim \loi{G}(a_l, b_l), &q &\sim \loi{N}(1, \sigma_q^2), &r &\sim \loi{G}(a_r, b_r), \label{priorlqr}
\end{align}
where \(a_l, b_l\), \(a_r, b_r\) and \(\sigma_q^2\) are fixed positive real numbers such that the prior distribution on \(l\), \(q\) and \(r\) are vague. These prior distributions are chosen because of their conjugacy properties (that we took advantage of to design the MCMC algorithms for our applications). The vagueness requirement that we impose on these priors is motivated by the fact that we want to keep as general a framework as possible without having to tweak each and every prior coefficient for different applications.

In the end, the informative prior is built as follows:
\begin{align}
\pi_\cB(\theta_\cB, k, l, q, r) &\propto \pi(\theta_\cB|k, l)\pi(k | q, r)\pi(l)\pi(q)\pi(r)\label{hierarchicalprior}
\end{align}
where the densities appearing on the right hand side are specified in \eqref{prioreta}, \eqref{priork} and \eqref{priorlqr}.

From the two first moments of the prior distribution $\pi_\cB$, we evaluate the prior information brought by the historical sample $y^\cA$. Firstly we show that the prior distribution $\pi_\cB$ is centered on the estimation of the parameter on $y^\cA$, which should be close to $\theta_\cA$ according to the asymptotic properties of Bayes estimates. 
\begin{property}
 The prior distribution admits a moment of order 1 equal to $ \widehat{\mu}^{\cA}$.
 \end{property}
\begin{proof}
 Indeed, we have 
 \begin{align*}
 \esp(\theta_\cB) &= \esp ( \esp(\theta_\cB | K, l^{-1}) ) = \esp( K \widehat{\mu}^{\cA}) \\ 
 &= \esp(K) \widehat{\mu}^{\cA} = \esp( \esp(K | q,r ) ) \widehat{\mu}^{\cA} = \esp(q) I_d \widehat{\mu}^{\cA} \\ 
 &= \widehat{\mu}^{\cA}. 
 \end{align*}
 \end{proof}
Secondly, from the correlation matrix of $\pi_\cB$, we show that we transfer informations from the historical sample to the dependence between the components of parameter $\theta_\cB$
\begin{property}
Assume that $a_l >1$ and $a_r>1$. Under suitable conditions of regularity,
\begin{itemize}
 \item the correlation of the prior distribution is equivalent to the correlation of the posterior distribution $\pi_\cA( \cdot | y^\cA)$ as $N_\cA$ goes to infinity.
 \item for a fixed $N_\cA$ the correlation of the prior distribution is also equivalent to the correlation of the posterior distribution $\pi_\cA(\cdot | y^\cA)$ as $\sigma^2_q$ and $\esp(r^{-1})$ go to 0.
 \end{itemize}
 \end{property}

\begin{proof}
First notice that the condition on the coefficients $a_l >1$ and $a_r>1$ ensures that the first moment of both $l^{-1}$ and $r^{-1}$ are finite.
This implies that the variance of the prior distribution is finite 

Recall now that if $(X,Y)$ are two random variables in $L^2$ we have the following equality
$$ \var(X) = \var(\esp(X|Y)) + \esp(\var(X|Y)). $$ 

We then get
\begin{align*}
 \var(\theta_\cB) &= \var( \esp(\theta_\cB | K, l^{-1}) ) + \esp ( \var(\theta_\cB | K, l^{-1}) ) \\ 
&= \var(K \widehat{\mu}^{\cA} ) + \esp(l^{-1} ) \hat\Sigma^{\cA} \\ 
&= \text{diag}( \widehat{\mu}^{\cA} ) \var(k) \text{diag}(\widehat{\mu}^{\cA} ) + \frac{b_l}{a_l-1} \hat\Sigma^{\cA}.
\end{align*}
Using the same equality, we can evaluate the variance of $k$ as
\begin{align*}
 \var(k) &= \var( \esp(k |q,r^{-1}) ) + \esp ( \var(k | q,r^{-1}) ) \\
&= \var(q \begin{pmatrix}
 1\\ \vdots\\ 1
\end{pmatrix} ) + \esp(r^{-1}) I_d \\ 
&= \sigma^2_q \begin{pmatrix}
 1 &\cdots & 1 \\ \vdots & \ddots & \vdots\\ 1 &\cdots & 1 
\end{pmatrix} + \frac{b_r}{a_r-1} I_d.
\end{align*}
We obtain
\begin{align*}
 \var(\theta_\cB) &= 
 \text{diag}( \widehat{\mu}^{\cA} ) \left( \sigma^2_q \begin{pmatrix}
 1 &\cdots & 1 \\ \vdots & \ddots & \vdots\\ 1 &\cdots & 1 
\end{pmatrix} + \frac{b_r}{a_r-1} I_d \right) \text{diag}(\widehat{\mu}^{\cA} ) + \frac{b_l}{a_l-1} \hat\Sigma^{\cA} \\ 
&= \frac{b_l}{a_l-1} \widehat\Sigma^{\cA} + \frac{b_r}{a_r-1} \text{diag}(\widehat{\mu}^{\cA} )^2 
+\sigma^2_q\, \widehat{\mu}^{\cA} {}^{t}\!\widehat{\mu}^{\cA}.
\end{align*}
and the element of the correlation matrix in i th row and j th column ($i\not=j$) is thus
$$
\frac{ \frac{b_l}{a_l-1} \widehat\Sigma^{\cA}(i,j) 
+\sigma^2_q\, \widehat{\mu}^{\cA}(i) \widehat{\mu}^{\cA} (j) } { \left(\frac{b_l}{a_l-1} \widehat\Sigma^{\cA}(i,i) +( \frac{b_r}{a_r-1} +\sigma^2_q) \widehat{\mu}^{\cA} (i) ^2 \right)^{1/2}\left(\frac{b_l}{a_l-1} \widehat\Sigma^{\cA}(j,j) + (\frac{b_r}{a_r-1} +\sigma^2_q )\widehat{\mu}^{\cA} (j) ^2 \right)^{1/2 } }.
$$

Under suitable conditions of regularity, $\widehat{\mu}^{\cA}$ converges to $\theta_\cA$ and $\frac{1}{N_\cA} \hat\Sigma^{\cA} $ converges to the Fisher information $I(\theta_\cA)$ as $N_\cA$ tends to infinity, which ends the proof of the first assertion.
The second assertion is trivial when you recall that $\esp(r^{-1}) = \frac{b_r}{a_r-1}$.
\end{proof}

\section{Basic examples}\label{sec:toyexamples}
In this Section, we describe how to apply our methodology to three different and common examples. We first deal with both a regression model and an auto-regressive model, before addressing a hierarchical Poisson model.
\subsection{Regression model}\label{exampleregression}
We will consider a Gaussian regression model i.e. for each \(t=1,\ldots,n\)
\begin{align}
y_t = f_t(\theta) + \epsilon_t \label{eq.modeleft}
\end{align}
where \(\epsilon_1, \ldots, \epsilon_N\) are assumed independent and identically distributed with common distribution \(\loi{N}(0, \sigma^2)\).

We will use the notation \(f(\theta) = (f_1(\theta),\ldots,f_N(\theta))\) for short, where \(\theta\) designate the parameters of interest. With these notations, since \(\eta=(\theta, \sigma^2)\), the likelihood of the model considered is
\begin{align}
L(y|\eta)=\left(\sqrt{2\pi}\sigma\right)^{-N}\exp\left(-\frac{1}{2}\|y-f(\theta)\|^2\right). \label{eq.likelihood}
\end{align}

For \(\sigma^2\) we use a non-informative prior (we chose \(\pi(\sigma^2) \propto \sigma^{-2}\)) because we do not want to make any kind of assumptions about the noise around both datasets. This prior is non-informative in the sense that it matches Jeffreys' prior distribution on \(\sigma^2\) for a Gaussian linear regression. For the parameter $\theta$ we use the informative hierarchical prior as described in Section \ref{Pgauss}

On the sample $y^\cA$ we use the following non-informative prior 
\begin{align}\label{Noninforegression}
\pi_\cA(\eta_\cA) =\pi_\cA(\theta_\cA,\sigma_\cA^{2}) &\propto \sigma_\cA^{-2}.
\end{align}
This prior is non-informative in the sense that it matches Jeffreys' prior distribution on \(\sigma_\cA^2\) for a Gaussian linear regression \cite[see][on page 73]{MarinRobert}. It leads to the following posterior distribution
\begin{align}
{\pi_\cA(\eta_\cA | y)} &\propto {L(y^\cA |\eta_\cA)} {\pi_\cA(\eta_\cA)} \label{noninfopost}\\
&\propto \sigma_\cA^{-N_\cA-2} \exp\left(-\frac{1}{2} \sigma_\cA^{-2} \|y - f(\theta_\cA)\|_2^2\right). \nonumber
\end{align}

We consider a polynomial regression model of order \(p\) \((p\geq0)\) on the interval \(\interff{-1}{1}\), meaning that observations from this model can be written for each \(t=1,\ldots,n\)
\begin{align}
y_t = \sum_{i=0}^p \theta_i x_{t}^p + \epsilon_n \label{eq:modelpoly}
\end{align}
where \(\epsilon_n\) are i.i.d. with Gaussian distribution \(\loi{N}(0, \sigma^2)\), and where \(x_1,\ldots,x_n\) are equally spaced over \(\interff{-1}{1}\).

Here again, we only show some of the results we obtained in various settings. Our goal here is to show, for a fixed sample size, the robustness of our method to non-exact similarity between the two datasets

\paragraph{First situation}
We simulate \(N_{\cA}=200\) observations \(y^{\cA}\) from model \eqref{eq:modelpoly} with \(p=4\) and coefficients \[\theta^{\cA}=(2, -1, 3, 1, 2)\] and $N_\cB = 10$  observations \(y^{\cB}\) with coefficients \[\theta^{\cB}= \rho\theta^{\cA},\] where \(\rho\) will be chosen later, using \(\sigma^2=1\) for the variance of the noise for \(y^\cA\) and \(\sigma^2=4\) for \(y^\cB\). We then estimate the coefficients \(\theta^{\cB}\) from the  small sample \(y^{\cB}\) available using the Gaussian informative prior we presented above. To assess the performance of our method, we compare and aggregate the results over 100 repetitions (sampling new observations for each repetition) with those obtained from a non-informative vague prior and those coming from MLE estimations. 
\paragraph{Second situation}
The second situation is different from the first in that \(\theta^{\cB}\) is not proportional to \(\theta^{\cA}\) anymore. Here we chose
\begin{align*}
\theta_0^{\cB} &= \rho \theta_0^{\cA}\\
\theta_1^{\cB} &= \rho \theta_1^{\cA}\\
\theta_2^{\cB} &= \theta_2^{\cA}\\
\theta_3^{\cB} &= \theta_3^{\cA}\\
\theta_4^{\cB} &= \theta_4^{\cA}\\
\end{align*}
and the rest of the setup is identical.

\begin{figure}[hbtp]%
\begin{center}
\includegraphics[width=.90\columnwidth]{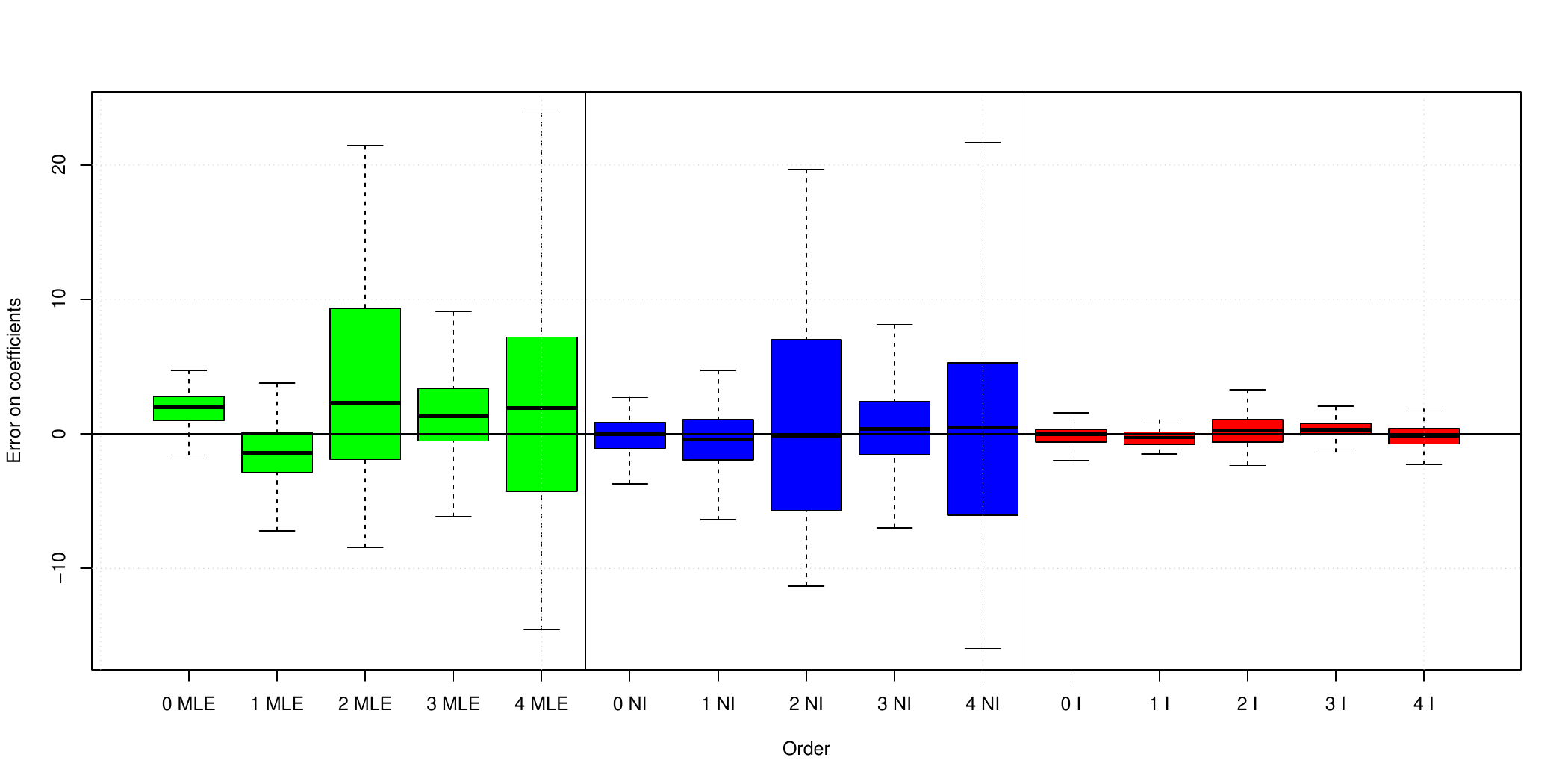}

\includegraphics[width=.90\columnwidth]{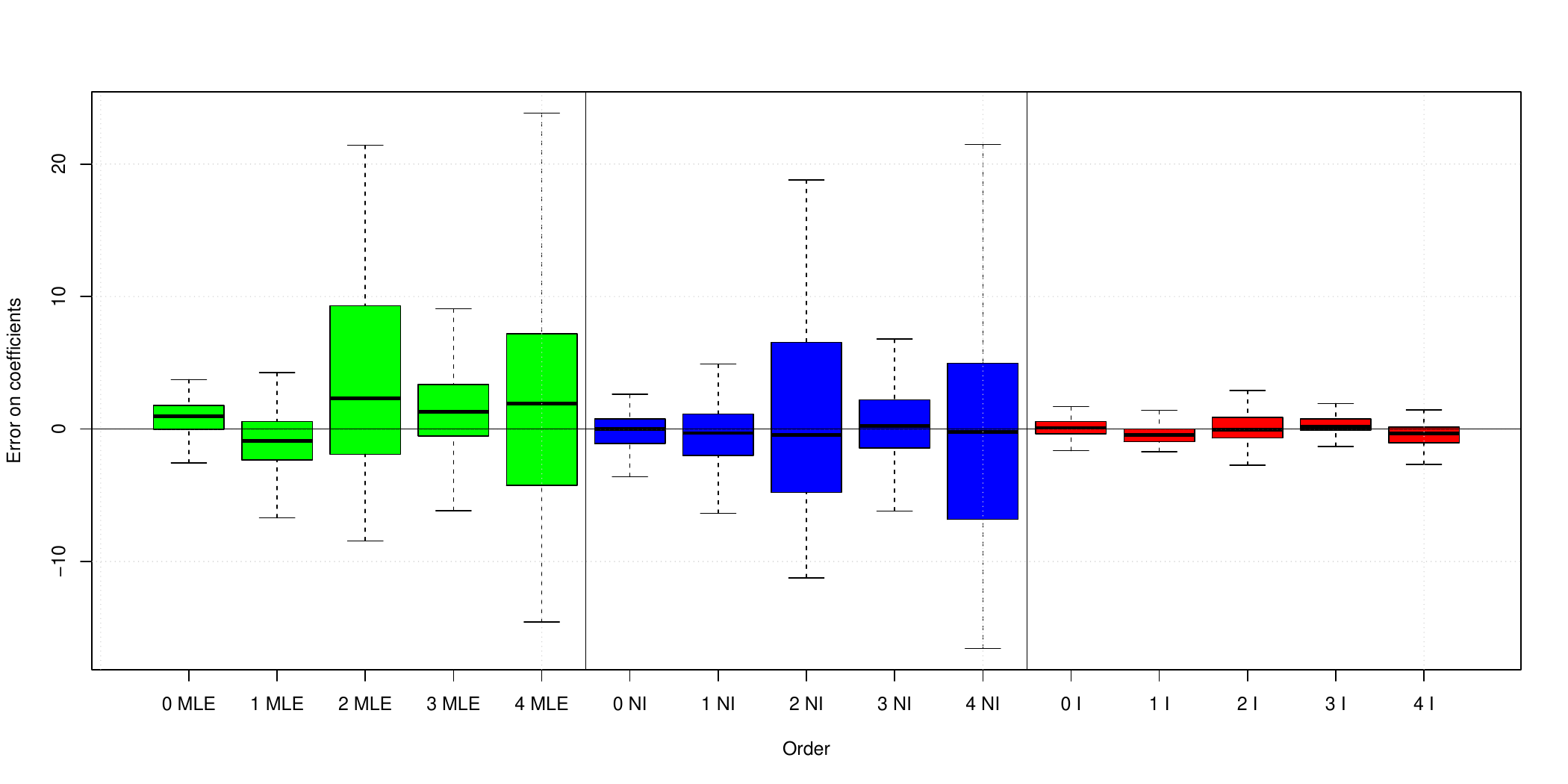}

\includegraphics[width=.90\columnwidth]{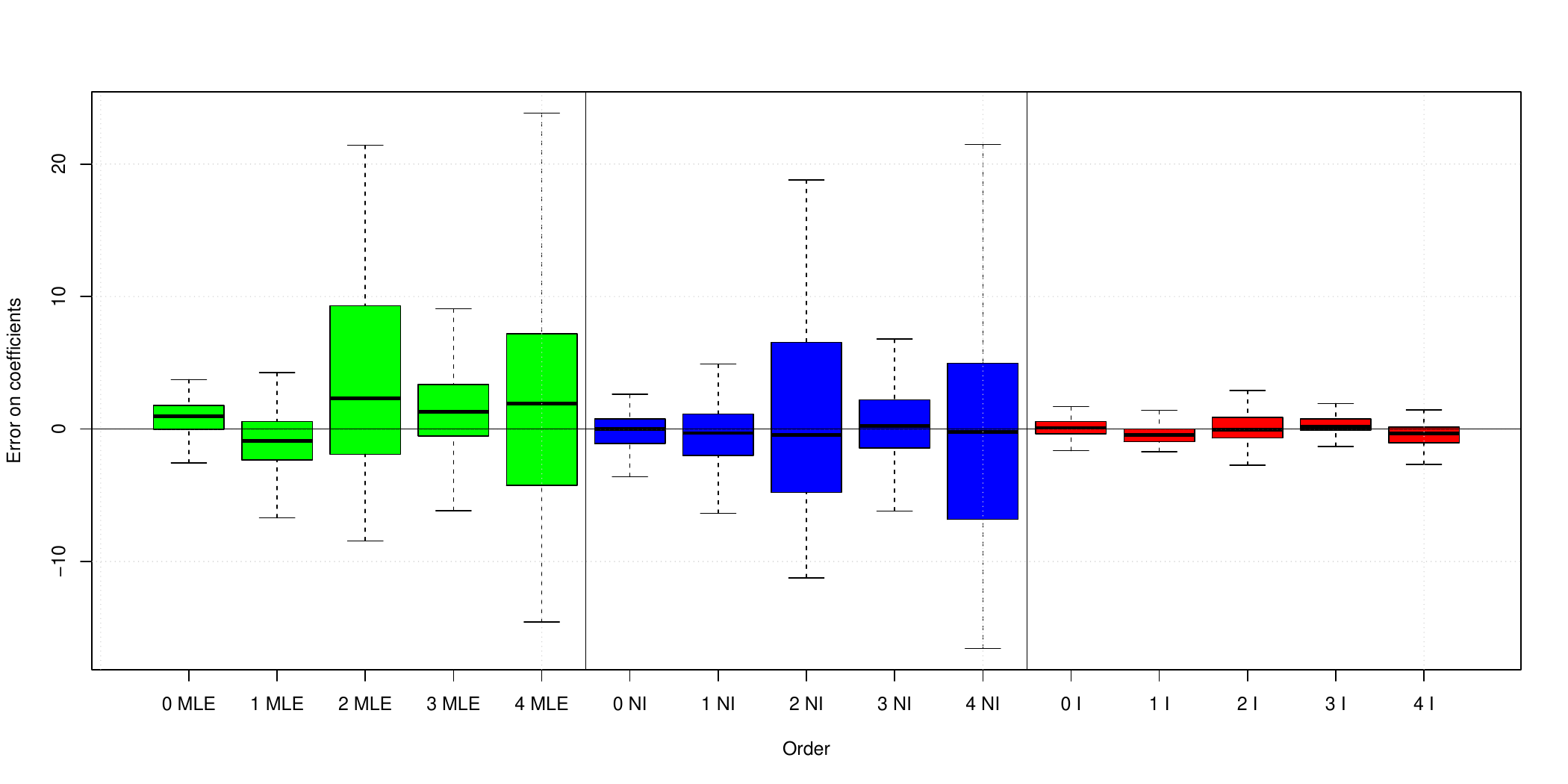}
\caption{Estimation errors on the four coefficients of the polynomial regression model \eqref{eq:modelpoly}, using MLE (in green), non-informative Bayesian (NI, in blue) and informative Bayesian (I, in red) methods, when \(\rho=1\) (top row), \(\rho=0.5\) in the first situation (middle row), \(\rho=0.5\) in the second situation (bottom row) }%
\label{fig:reg_theta}%
\end{center}
\end{figure}

\begin{figure}[hbtp]
\begin{center}
\includegraphics[width=.49\columnwidth]{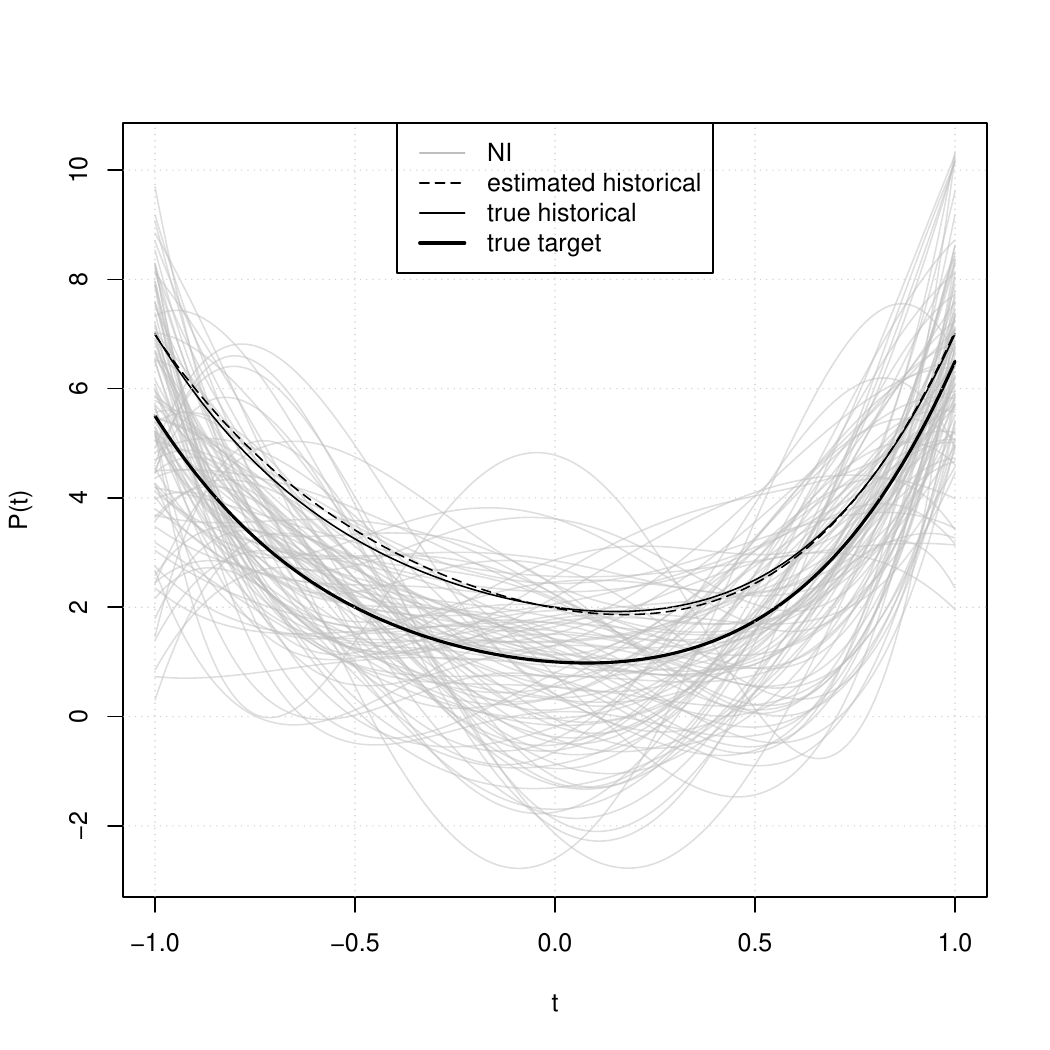}
\includegraphics[width=.49\columnwidth]{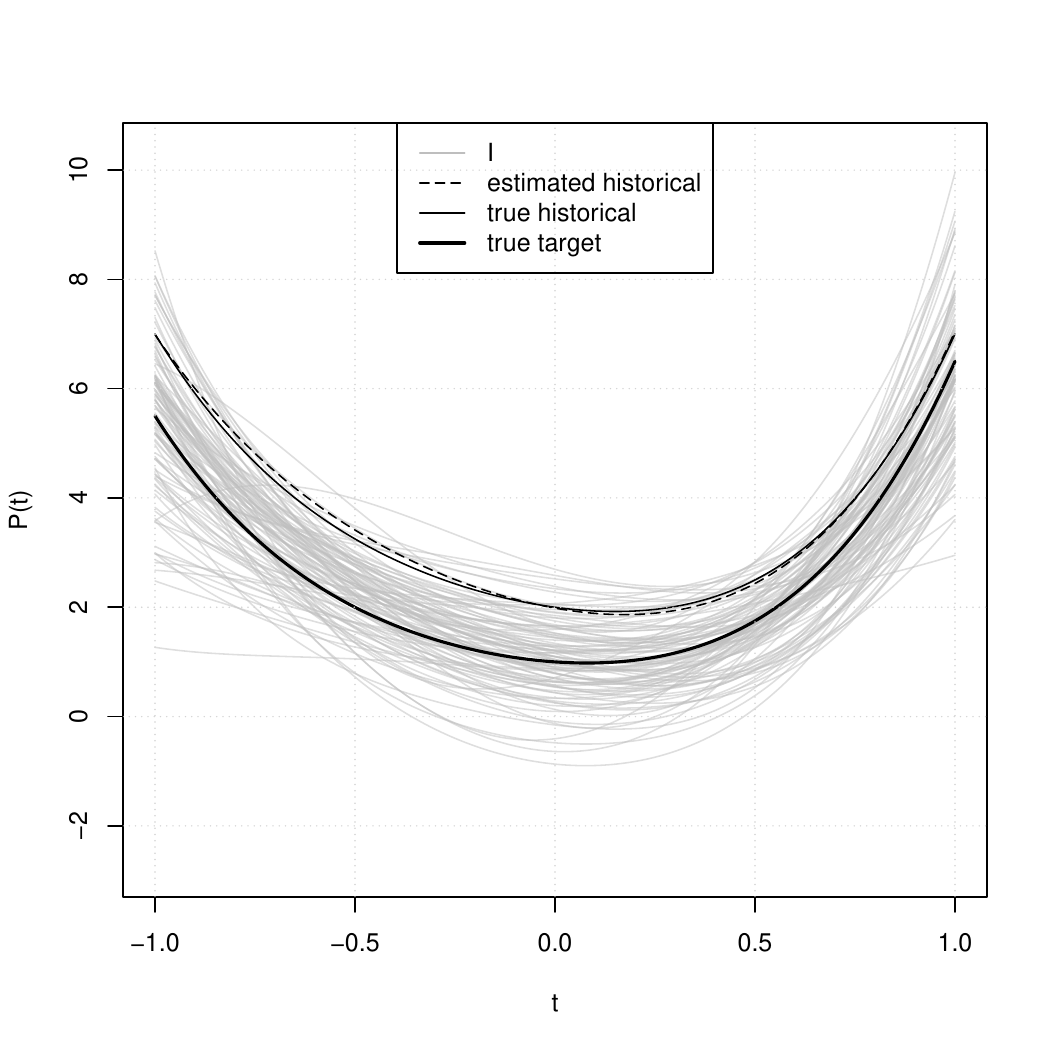}

\includegraphics[width=.49\columnwidth]{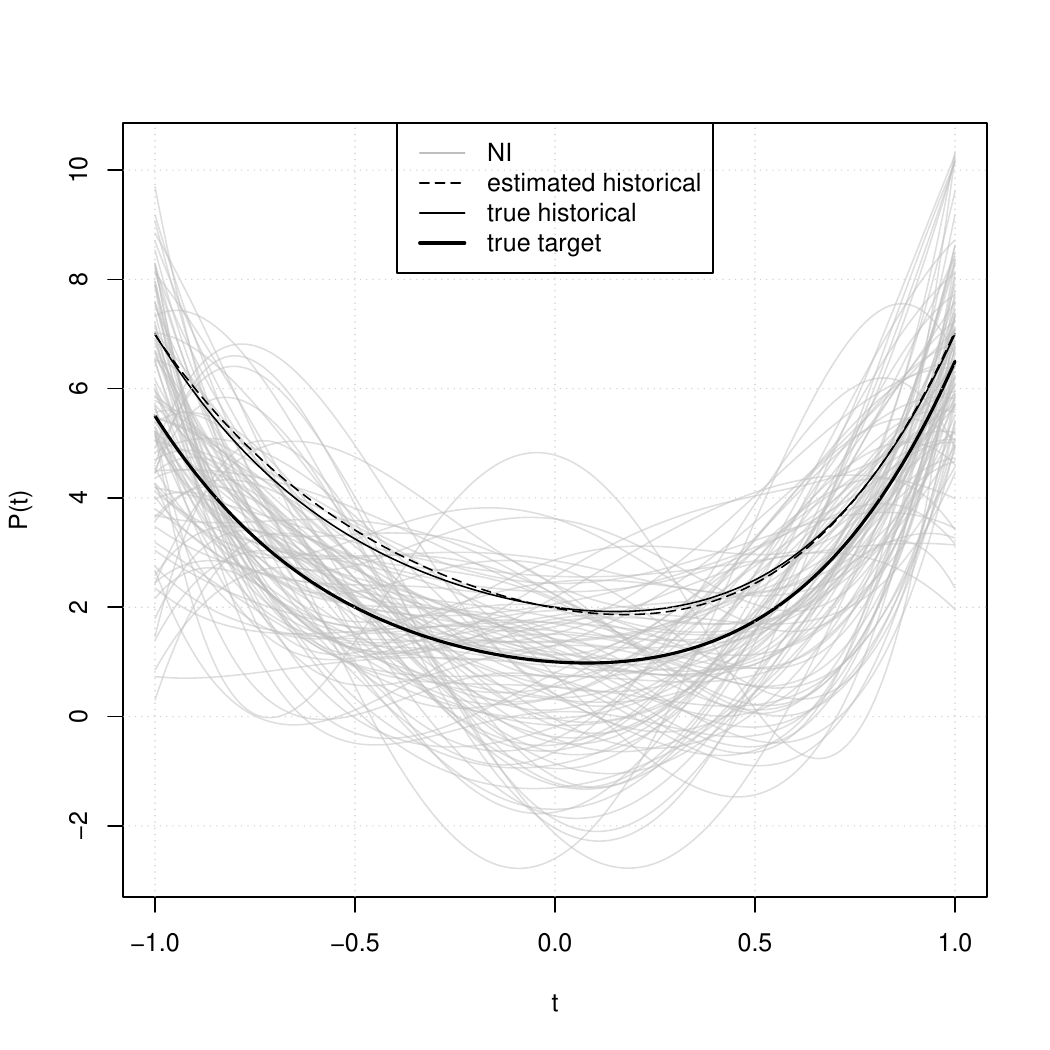}
\includegraphics[width=.49\columnwidth]{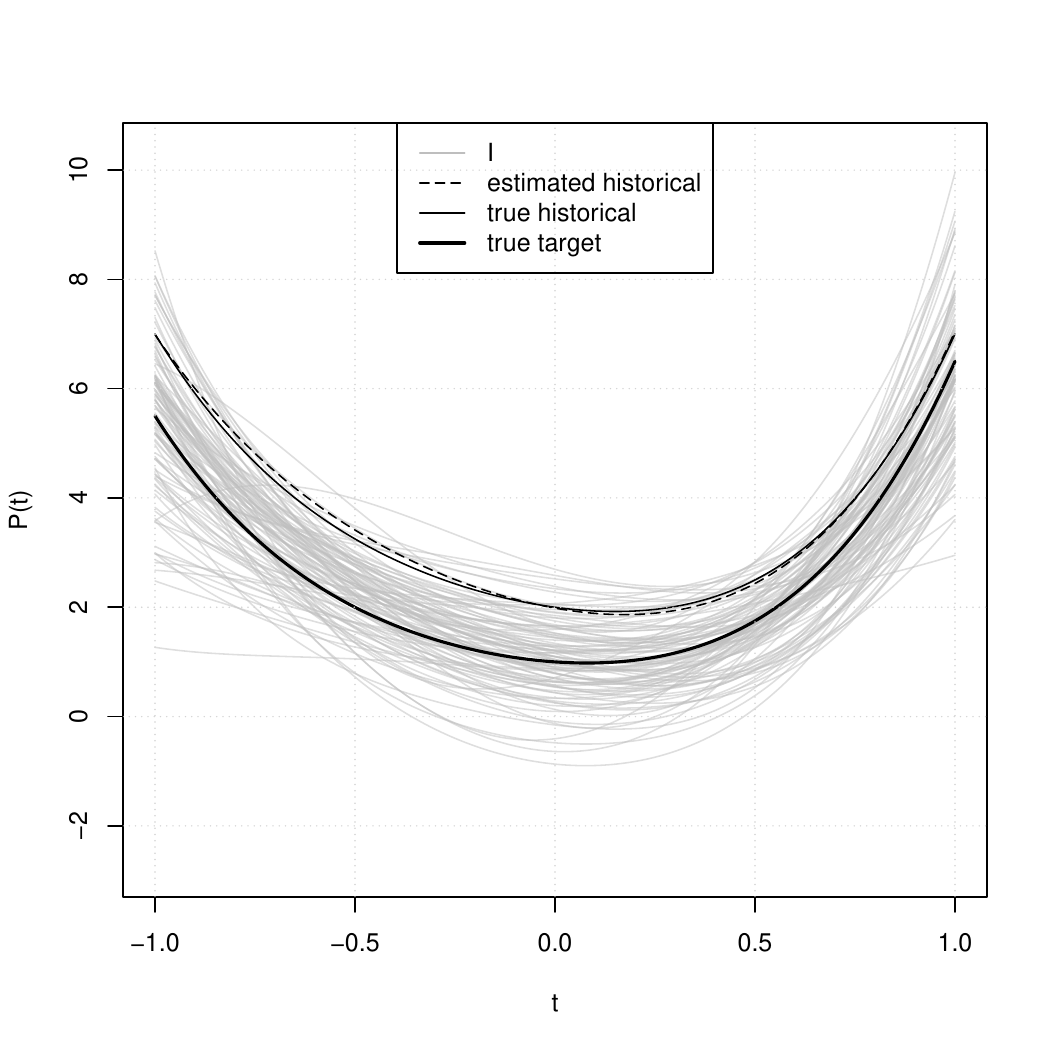}
\caption{Estimated polynomials using non-informative Bayesian prior (left) and informative Bayesian (right). The thin solid line represents the true polynomial corresponding to \(\theta^{\cA}\), the dashed line the estimated polynomial on observations \(y^{\cA}\), and the thick solid line represents the true polynomial corresponding to \(\theta^{\cB}\). Top row shows results for the first situation, while bottom row shows results for the second situation.}%
\label{fig:reg_5_poly}%
\end{center}
\end{figure}

\paragraph{Results.}
Figure \ref{fig:reg_theta} shows the errors of estimation for the four coefficients i.e. the true values minus the estimated values (the posterior mean for Bayesian models) when \(\rho\) varies from \(1\) to \(0.5\) for both situations and Figure \ref{fig:reg_5_poly} displays the actual estimated polynomials for both situations when \(\rho=0.5\).

Once again the benefits of using the informative prior over the non-informative prior stand out. Whether the similarity between the two parameters \(\theta^{\cB}\) and \(\theta^{\cA}\) is based upon proportionality or not the informative prior helps the estimation of all the coefficients in a very effective way.

As the sample size \(N_{\cB}\) decreases from 100 to 10, the MLE and the non-informative Bayesian method tend to perform rather poorly whereas the use of the informative Bayesian prior clearly benefits the estimation of the coefficients in this situation where observations \(y^{\cA}\) and \(y^{\cB}\) come from exactly the same model.

\subsection{Auto-regressive model}

We consider an AR(\(p\)) model \((p>0)\) with coefficients \(\theta\), and variance of noise \(\sigma^2\), i.e. denoting \(y_n\) its observation at time \(n\), for each \(n\)
\begin{align}
y_n = \sum_{i=1}^p \theta_i y_{n-i} + \epsilon_n \label{eq:modelAR}
\end{align}
where \(\epsilon_n\) are i.i.d. with Gaussian distribution \(\loi{N}(0, \sigma^2)\).

As in the previous example, we choose for $\theta$ the informative hierarchical prior described in Section \ref{Pgauss}, and for \(\sigma^2\) we take the non-informative prior proportional to \(\sigma^{-2}\) . On the historical sample $y^\cA$, we again choose the non informative prior \eqref{Noninforegression} \cite[see][on page 219]{MarinRobert}.

We tested our method on many different such models (with varying orders, coefficients, and variance of noise) and present below only the results obtained for an auto-regressive model of order 4 \((p=4)\) and show the effect of the sample size \(N_{\cB}\).

\begin{figure}[hbtp]%
\begin{center}
\includegraphics[width=.90\columnwidth]{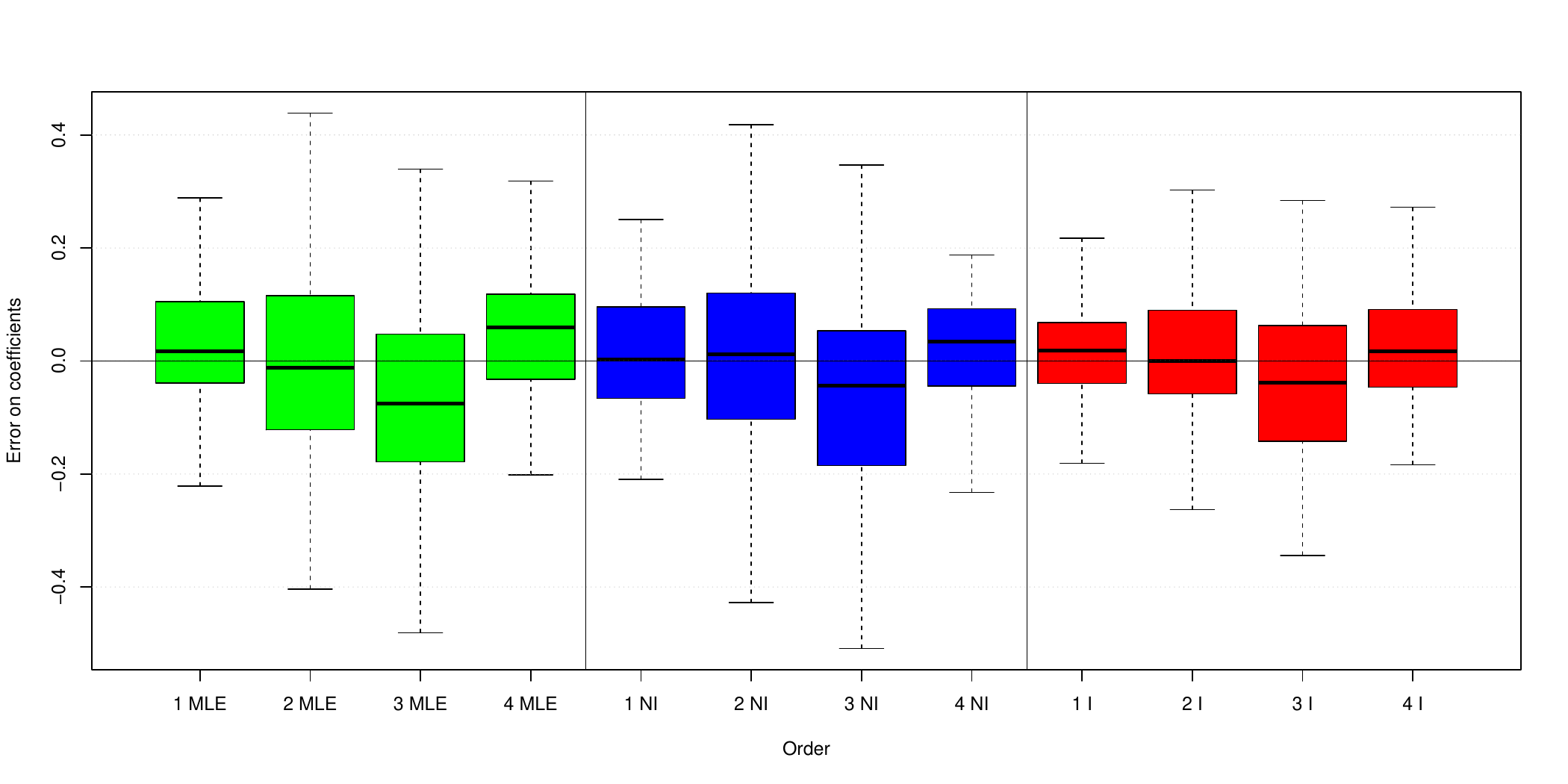}

\includegraphics[width=.90\columnwidth]{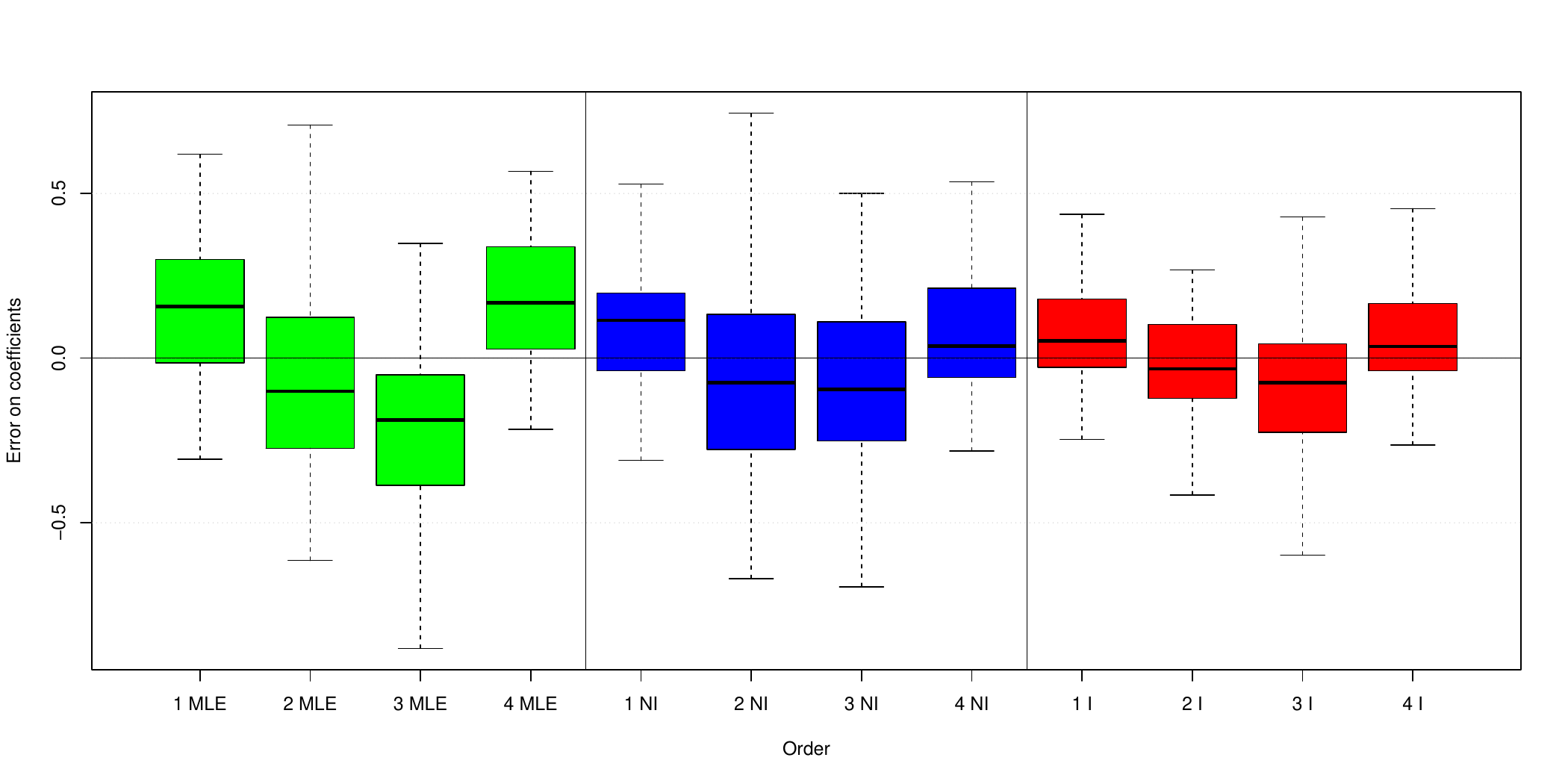}

\includegraphics[width=.90\columnwidth]{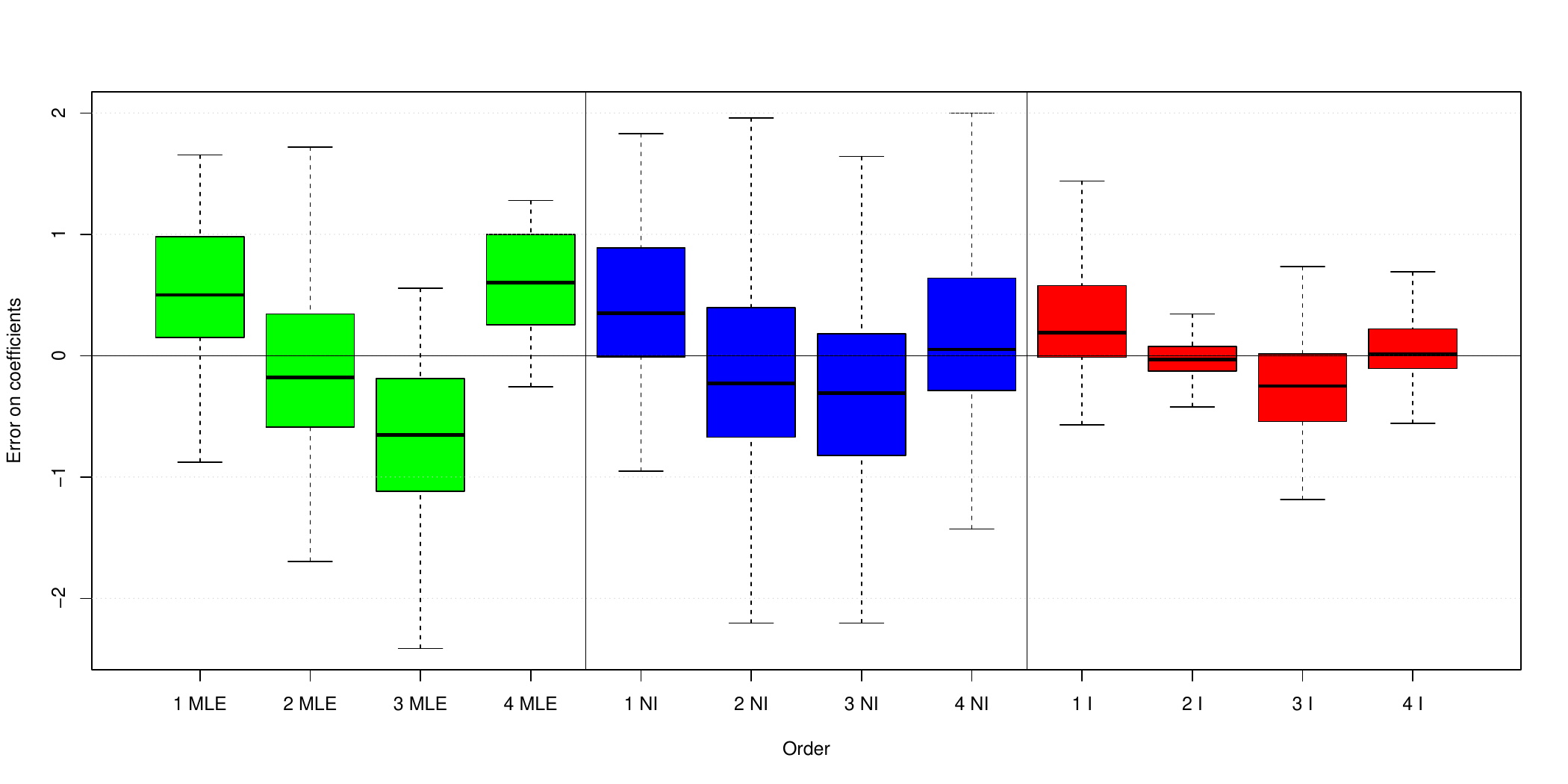}
\caption{Estimation errors on the four coefficients of the auto-regressive model \eqref{eq:modelAR}, using MLE (in green), non-informative Bayesian (NI, in blue) and informative Bayesian (I, in red) methods, for three different sample sizes : \(N_{\cB}=100\) (top row), \(N_{\cB}=30\) (middle row), \(N_{\cB}=10\) (bottom row) }%
\label{fig:AR_4_coef}%
\end{center}
\end{figure}

We simulate \(N_{\cA}=400\) observations \(y^{\cA}\) from model \eqref{eq:modelAR} with coefficients \[\theta^{\cA}=(1.7, -0.23, -0.833, 0.3528)\] (the corresponding polynomial having roots \(-0.7, 0.7, 0.8\) and \(0.9\)), and also simulate \(N_{\cB}=100,30,10\) observations \(y^{\cB}\) from the same AR(4) model with coefficients \[\theta^{\cB}=\theta^{\cA},\] using in each case \(\sigma^2=1\) for the variance of the noise. We then estimate the coefficients \(\theta^{\cB}\) from the observations \(y^{\cB}\) available using the Bayesian prior we presented above. To assess the performance of our method, we compare and aggregate the results over 100 repetitions (sampling new observations for each repetition) with those obtained from a non-informative vague prior and those coming from MLE estimations. Figure \ref{fig:AR_4_coef} shows the errors of estimation for the four coefficients i.e. the true values minus the estimated values (the posterior mean for Bayesian models).

\subsection{Hierarchical Poisson model} 
\label{sec:hier-poiss-model}
We consider $y_1,\ldots,y_N$ independent random variables such that for each $j$, $y_j$ follows a Poisson distribution with parameter $\lambda_j$. We assign a Gamma distribution for the parameters $\lambda_i$ : 
\begin{align*}
 y_i &\sim \mathcal{P} (\lambda_i)\\ 
 \lambda_i &\sim \Gamma (a,b)
\end{align*}

When a large sample is available, we can choose a non informative distribution on the parameter $\theta = (a,b)$, for example the Jeffreys prior which is given by 
$$ \pi_J(a,b) = \frac 1b \sqrt{a (\log(\Gamma(a)) '' -1}$$ 
(see \cite{Yangberger98} ). When $a$ goes to zero or to infinity, 
$\sqrt{ a (\log(\Gamma(a)) '' -1}$ is equivalent to $ \frac{1}{\sqrt{a}}$
(see \cite{abramovitzstegun65} chapter 6). 
To simplify the computational aspects, we consider a prior distribution which behaves as the Jeffreys prior when $a$ tends to zero and to infinity. 
We assume that $a$ and $b$ are independent, the marginal distributions of $a$ and $b$ are Gamma distributions with parameters $(1/2, \epsilon)$ and $(\epsilon, \epsilon) $ respectively. The parameter $\epsilon$ is fixed small enough. 

\begin{figure}[hbtp]%
\begin{center}
\includegraphics[width=\textwidth,height=.5\textheight]{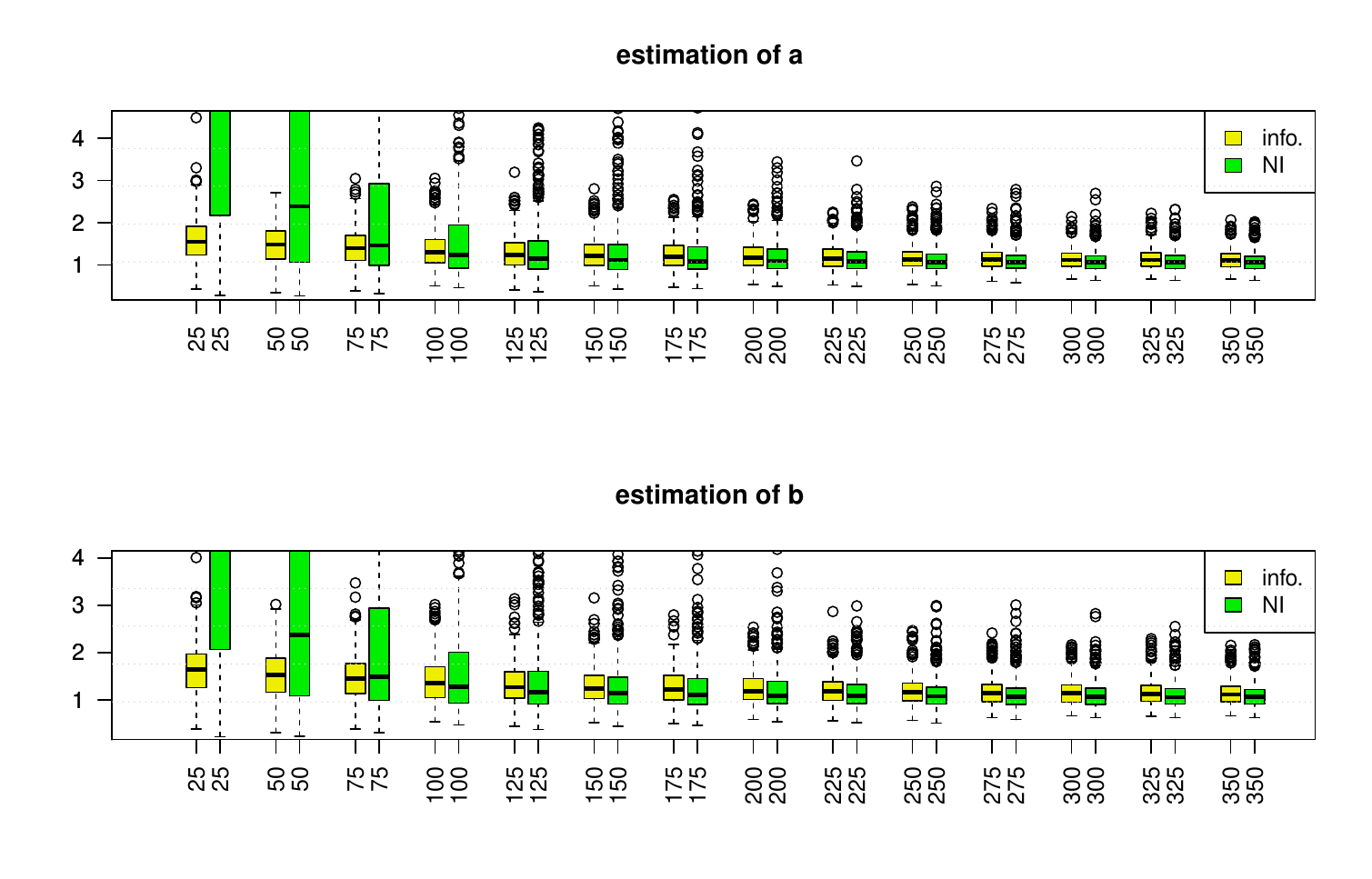}
\caption{Estimation of the hyperparameters $\theta_\cB = (a,b)$ using hierarchical informative prior (Info.) and non informative prior as function of sample sizes $N_{\cB} $. 
The true gamma parameters are $(a_\cA,b_\cA) =(a_\cB,b_\cB)=(1,1)$. The boxplot is constructed from 500 independent replications.}
\label{fig:pois2}%
\end{center}\end{figure}

For this choice of prior, Figure \ref{fig:pois2} illustrates that the Bayes estimates do not perform well for the small sample. 

In the context of Section \ref{sec:methodo}, there is a large sample $y^\cA$ sharing some similarities with the small 
sample of interest $y^\cB$. Therefore we can construct the prior distribution on $(a,b)$ using the sample $y^\cA$ 
based on either of the two following strategies
\begin{enumerate}
\item we assign on $(a,b)$ the hierarchical prior defined in \eqref{hierarchicalprior} with $\theta^\cB =(a,b)$. 
\item the parameter $(a,b)$ are fixed, we take the values of the estimates calculated on the large sample \(y^\cA\) (See \citet{albert2009Bayesian,Gelmanetal})
\end{enumerate}

\begin{figure}[hbtp]%
\begin{center}
\includegraphics[width=\textwidth,height=.5\textheight]{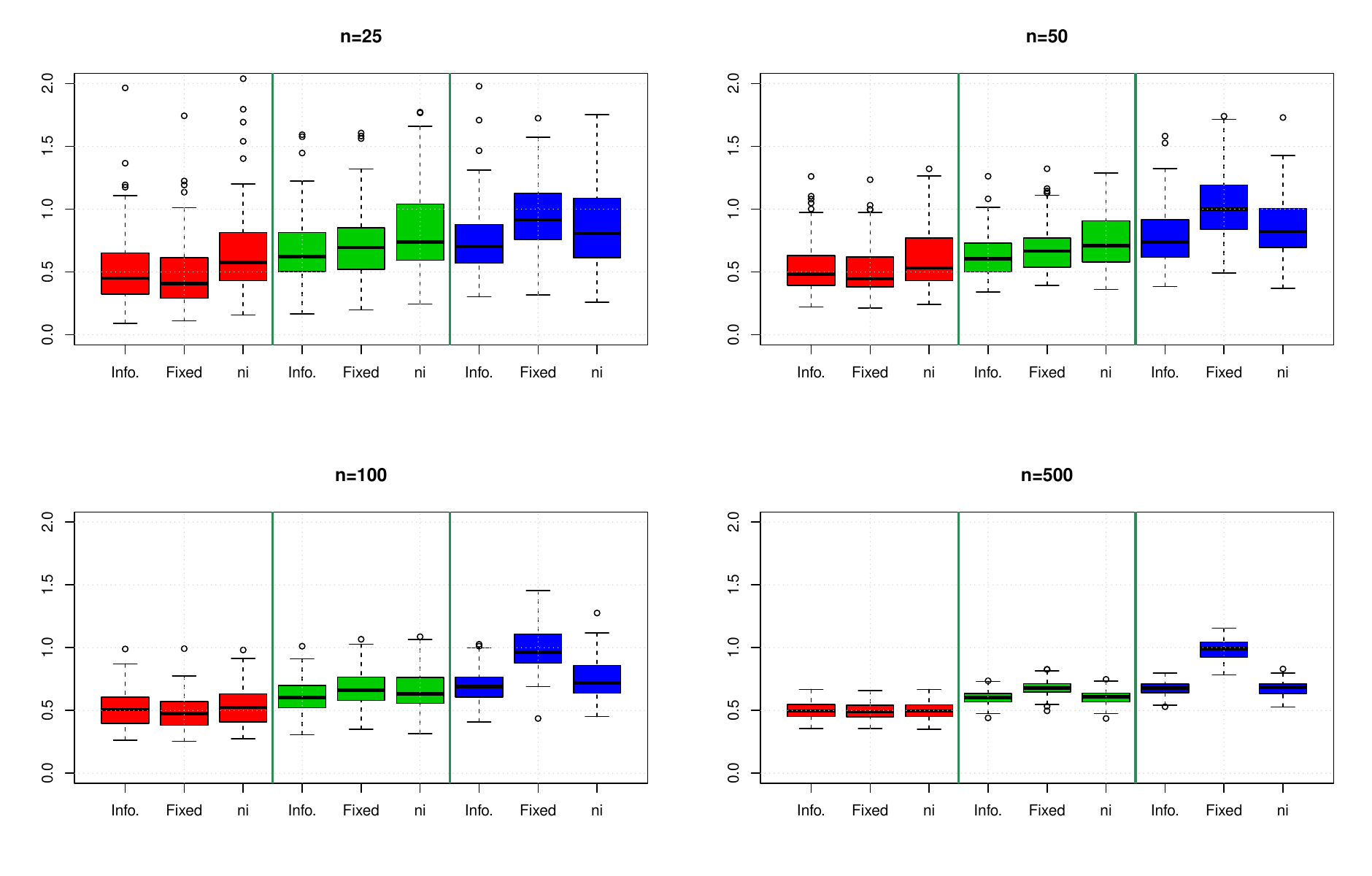}
\caption{Mean square error $E_n$ defined in \eqref{eq:err} for the three choice of prior : hierarchical prior (Info.), fixed hyperparameter (Fixed) and non-informative prior (ni). The parameters are $\theta_\cB =  (a_\cB,b_\cB)$: \( (1,1)\) (red), \((9/4,3/2)\) (green), \((4,2)\) (blue). The sample sizes are $N_\cB =25,\; 50,\; 100,\; 500$ and $N_\cA=5000$. The gamma parameters of $y^\cA$ are $\theta_\cA =  (a_\cA,b_\cA) =(1,1)$. The boxplot is constructed from 500 independent replications }
\label{fig:pois1}%
\end{center}
\end{figure}

Figure \ref{fig:pois2} shows the improvement brought by the informative hierarchical model upon non informative model. 
On independent replications, we estimate the mean square error on the estimation of $\theta_i$ 
\begin{equation}
E_N = \frac{1}{N} \sum_{i=1}^N | \lambda_i-\esp(\lambda_i| y_\cB)|^2. \label{eq:err}
\end{equation}
The results are given in Figure \ref{fig:pois1}. When the two populations have exactly the same parameters, it is not surprising to obtain better results using fixed parameters. However the simulation shows also that this strategy is not robust and leads to worse results than the non informative prior when the true parameters differ. 
The hierarchical prior on $(a,b)$ gives some very satisfying results in terms of quality of the estimation and robustness to variation of the parameters between the sample 
$y^\cA$ and $y^\cB$.

\clearpage

\section{Electricity load forecasting}\label{sec:electricity}%
Modelling and forecasting the electricity load on a day-to-day basis has long been a key activity for any company involved in the electricity industry. It is first and foremost needed to supply a fixed voltage at all ends of an electricity grid: the amount of electricity produced has to match the demand very closely at any given time and experts usually make use of short-term forecasts \cite[see][for example]{Cottet}).

Electricity load usually involves a large predictable component due to its very strong daily, weekly and yearly periodic behaviour and the exogenous variables most commonly used to forecast the electricity load are weather-based \cite[see][for an example]{Alzayer,TaylorBuizza}. For the French electricity load specifically, the importance of the temperature and cloud-cover is underlined in \cite{Menage} and \cite{Bruhns}. \cite{EngleGrangerRice,RamanathanEngle} also present other weather-related models.

Some authors work with univariate time series models: \cite{Taylor} builds a double seasonal exponential smoothing for the British electricity load, and \cite{Cugliari} treats the load curve as a functional-valued autoregressive Hilbertian process to use a non-parametric approach relying on the wavelet transform. Other authors \cite[see][for example]{SoaresMedeiros,RamanathanEngle,Dordonnat} choose multiple-equation models where the various hours of the day share the same equation, while the associated parameters differ.

Bayesian methods have also been used in the past to solve electricity load forecasting problems. Amongst them, \cite{Minka} deals with Bayesian linear regression and \cite{HarrisonStevens} focus on dynamic linear models from the Kalman filter point of view and \cite{Smith} choose a Bayesian semi-parametric regression model. To the best of our knowledge though, no previous attempt has been made in the literature to help improve the forecasting of an intraday electricity load model on a short dataset, based upon another dataset using a hierarchical Bayesian model.

\subsection{Regression model}
In the next few paragraphs, we specify the regression model that we use to model the electricity load.

For each of the 48 instants of the day (each instant lasts 30 minutes, starting from 00:00\textsc{am}), the non-linear regression model that we consider in this paper, first described in \cite{Bruhns}, is made of three components and is usually formulated as follows: for \(t = 1,\ldots,N\),
\begin{align}
y_t &= x^{(1)}_t x^{(2)}_t + x^{(3)}_t + \epsilon_t \label{eventail}\\
x^{(1)}_t &= \sum_{j=1}^{d_{11}} \left[ z_j^{\cos} \cos \left(\frac{2j\pi}{365.25} t\right) + z_j^{\sin} \sin \left(\frac{2j\pi}{365.25} t\right) \right] + \sum_{j=1}^{d_{12}} \omega_j \one_{\Omega_j}(t), \nonumber\\
x^{(2)}_t &= \sum_{j=1}^{d_2} \psi_j \one_{\Psi_j}(t), \nonumber\\
x^{(3)}_t &= g (T_t - u)\one_{\interfo{T_t}{+\infty}}(u), \nonumber
\end{align}
where \(y_t\) is the load of day \(t\) and where \(\epsilon_1, \ldots, \epsilon_N\) are assumed independent and identically distributed with common distribution \(\loi{N}(0, \sigma^2)\).

\begin{figure}[tbp]
 \centering
 \includegraphics[width=.495\textwidth]{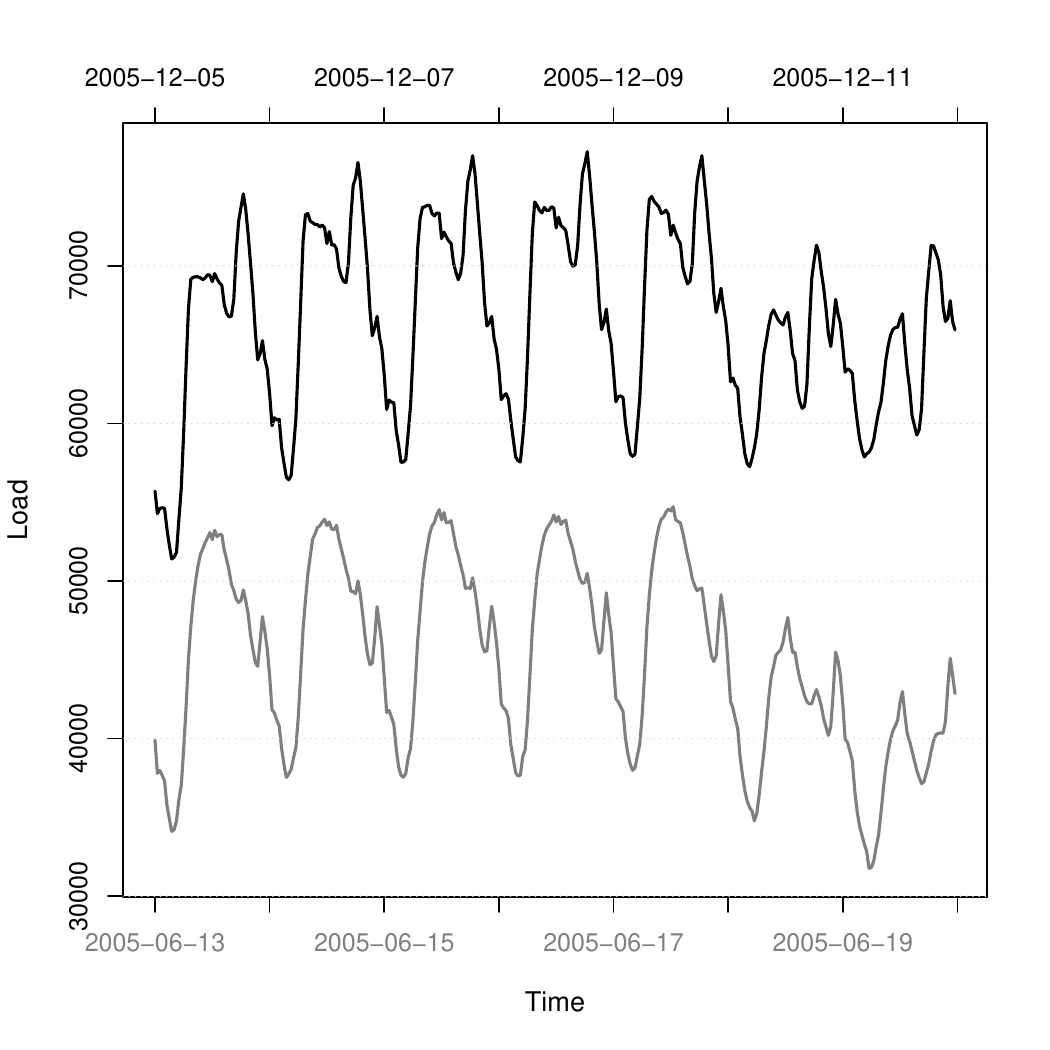}
 \includegraphics[width=.495\textwidth]{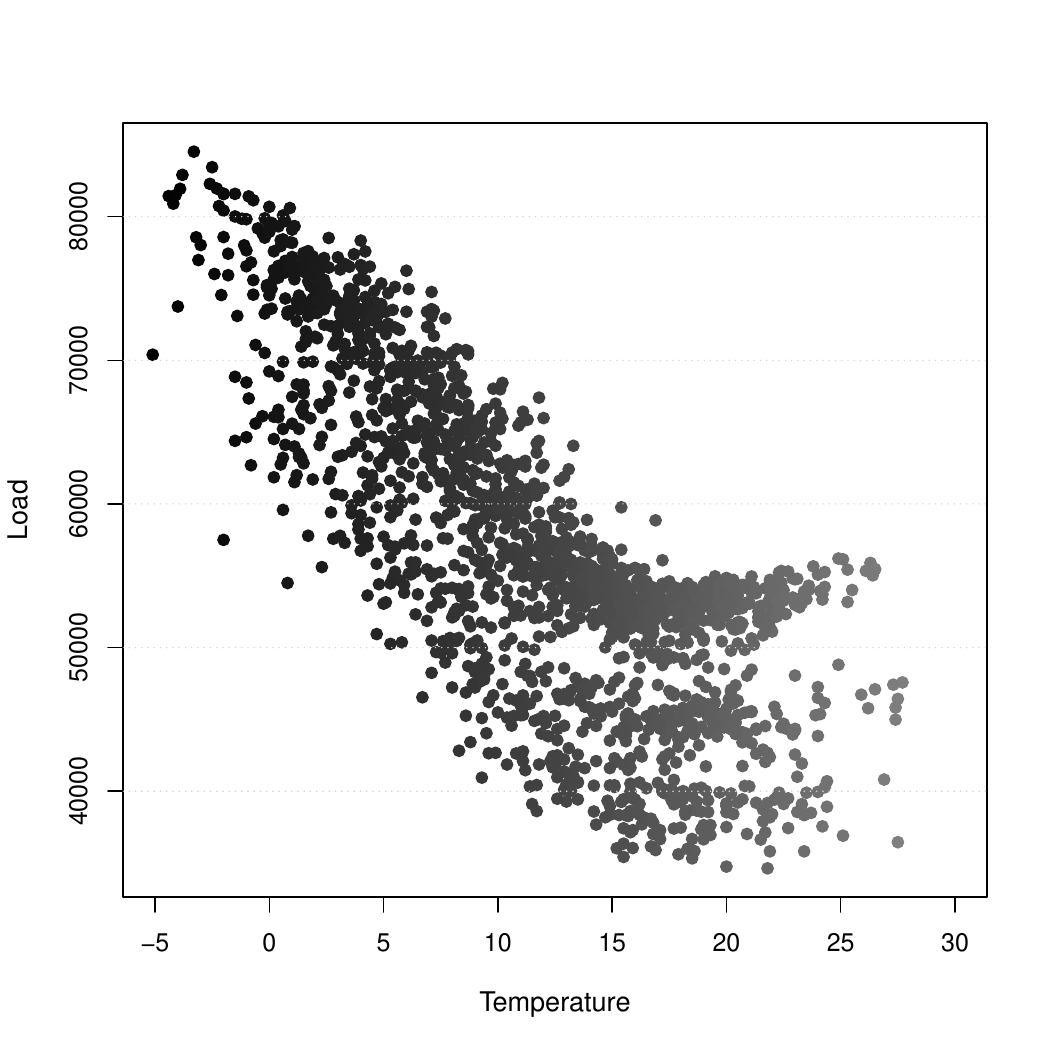}
 \caption{Left~: French Electricity load from 13/06/2005 to 29/06/2005 (in grey) and from 05/12/2005 to 11/12/2005 (in black). The load is expressed in MW. Notice the daily patterns of the electricity load are not the same during summer and winter. Right~: French Electricity load at 10:00 over 5 years against temperatures. The load seems to increase linearly with the temperature below a certain threshold.}
 \label{fig:illuconso}
\end{figure}

The \(x^{(1)}\) component is meant to account for the average seasonal behaviour of the electricity load, with a truncated Fourier series (whose coefficients are \(z_j^{\cos} \in \R\) and \(z_j^{\sin} \in \R\)) and gaps (parameters \(\omega_j \in \R\)) which represent the average levels of electricity load over predetermined periods given by a partition \(( \Omega_j)_{j\in\{1,\ldots,d_{12}\}}\) of the calendar. This partition usually specifies holidays, or the period of time when daylight saving time is in effect i.e. major breaks in the electricity consumption behaviour. The left part of Figure \ref{fig:illuconso} shows the typical behaviour of the electricity load over two different periods of time (summer vs. winter).

The \(x^{(2)}\) component allows for day-to-day adjustments of the seasonal behaviour \(x^{(1)}\) through shapes (parameters \(\psi_j\)) that depends on the so-called days' types which are given by a second partition \((\Psi_j)_{j\in\{1,\ldots,d_2\}}\) of the calendar. This partition usually separates weekdays from weekends, and bank holidays. The differences between two different daytypes are visible on the left part of Figure \ref{fig:illuconso} too. For obvious identifiability reasons, the vector \(\psi\) is restricted to the positive quadrant of the \(\|\cdot\|_1\)-unit sphere in \(\R^{d_2}\), that we denote \[S_+^{d_2}(0, 1) = \left\{ \psi \in (\R_+)^{d_2} ;\; \|\psi\|_1 = 1 \right\}. \]

The \(x^{(3)}\) component represents the non-linear heating effect that links the electricity load to the temperature \cite[see][for a general presentation of non-linear models]{SeberWild}, with the help of 2 parameters. The heating threshold \(u\in\interff{\underline{u}}{\overline{u}}\) corresponds to the temperature above which the heating effect is considered null and is usually estimated to be roughly around 15$^\circ$C. The heating effect is supposed to be linear for temperatures below the threshold and null for temperatures above. Note that the threshold is sought within the range of the observed temperatures, i.e. \(u\in\interff{\underline{u}}{\overline{u}}\) with \[ \min_{t=1,\ldots,N} T_t < \underline{u} < \overline{u} < \max_{t=1,\ldots,N} T_t. \] The heating gradient \(\gamma \in \R^*\) where \(\R^* = \R \backslash \{0\}\) represents the intensity of the heating effect, i.e. the slope (assumed to be non-zero) of the linear part that can be observed on the right part of Figure \ref{fig:illuconso}.

\subsection{Priors}
For the sake of both clarity and completeness, we present the full Bayesian model in use for all our applications in the lines below. Let us denote here again \(\eta = (\theta,\sigma^2)\) where \(\theta\) designates the parameters of interest in the electricity load model (i.e. \(z_j, \omega_j, \psi_j, g, \text{ and } u\)).

We opted for the Gaussian strategy (see Section \ref{Pgauss}) for the regression model as was described in Section \ref{exampleregression}, meaning that the non-informative prior distribution \(\pi_\cA\) used with the historical sample \(y^\cA\) is
\begin{align*}
\pi_\cA(\eta_\cA) &= \pi_\cA(\theta_\cA,\sigma_\cA^2) \propto \sigma_\cA^{-2}.
\end{align*}

The informative prior \(\pi_\cB\) used with the small sample \(y^\cB\) is defined hierarchically and can be summarised as
\begin{align*}
\pi_\cB(\theta_\cB, \sigma_{\cB}^2, k, l, q, r) &\propto \pi(\theta_\cB|k, l)\pi(k | q, r)\pi(l)\pi(q)\pi(r)\pi(\sigma^2)
\end{align*}
with:
\begin{align*}
\theta_\cB | k,l &\sim \loi{N}(K\widehat{\mu}^{\cA}, l^{-1}\widehat{\Sigma}^{\cA}) &  q &\sim \loi{N}(1, \sigma_q^2)\\
k|q,r &\sim \loi{N}(q(1,\ldots,1)^\prime, r^{-1} I_d) & r &\sim \loi{G}(a_r, b_r) \\
l &\sim \loi{G}(a_l, b_l) & \pi(\sigma^2) &\propto \sigma^{-2},
\end{align*}
where \(K=\text{diag}(k)\), where \(\widehat{\mu}^{\cA}\) and \(\widehat{\Sigma}^{\cA}\) are the posterior mean and posterior variance of the parameter \(\theta_{\cA}\) and where \(a_l, b_l\), \(a_r, b_r\) and \(\sigma_q^2\) are fixed positive real numbers such that the prior distribution on \(l\), \(q\) and \(r\) are vague.

\begin{remark}
Using either the non-informative prior distribution or the informative prior distribution with the model \eqref{eventail} leads to a proper posterior distribution. The proofs, as well as the MCMC algorithms we designed to derive numerical estimators, are pushed back into the Appendix (see Section \ref{sec:appendix}).
\end{remark}

\subsection{Datasets}\label{sec:applications}
The historical dataset \(y^\cA\) needed for the construction of the informative prior corresponds to a specific population in France frequently referred to as ``non-metered'' because their electricity consumption is not directly observed by EDF but instead derived as the difference between the overall electricity consumption and the consumption of the ``metered'' population. For this population the data ranged from 07/01/2004 to 07/31/2010.

We illustrate the benefits of choosing our informative prior to predict electricity load on two short datasets : the first dataset \(y^\cB\) corresponds to the ``non-metered'' population for ERDF, a wholly owned subsidiary of EDF that manages the public electricity network for 95\% of continental France. This population roughly covers the same people that \(y^\cA\) does, but not exactly. The population behind the second dataset \(y^{\cB\prime}\) corresponds to a subset of the population behind \(y^\cA\) and represents around 50\% of the total load of \(y^\cA\).

\subsection{Benchmark against standard methods}\label{subsec:applibenchmark}
For this application, we only consider the days for which no special tariffs are enforced: the so-called EJP (``Effacement jour de pointe`` = peak tariff days) are removed from the dataset beforehand to ensure the signal studied is consistent throughout time. Bank holidays (including the day before and the day after to avoid any neighbourhood contamination effects), the summer holiday break (August) and the winter holiday break (late December) are also removed from the dataset for this first application, so that we may benchmark our method against others on a smooth and rather easy-going signal, and not put any one method at a disadvantage due to the signal's specificity. The temperature considered in the model is the average temperature over France for the period of study, and the cooling threshold is chosen to be 16\(^\circ\)C throughout the 48 instants of the day.

We benchmark our Bayesian method with informative prior against four alternative methods, comparing their predictions on dataset \(y^\cB\) in two configurations. Roughly speaking, for our first experiment (Case 1) we estimate the model for \(y^\cB\) over the period ranging from 12/01/2009 to 06/30/2010 and predict the next 30 days (same as the application presented Section \ref{subsec:applihyper}), while for our second experiment (Case 2) we estimate the model for \(y^\cB\) over the period ranging from 01/01/10 to 07/31/10 and predict the previous 30 days. We expect Case 1 to be the easy case and Case 2 to be the tough case, the signal being very smooth during summer and not so much during winter. The figures shown in Table \ref{tab.datasize} summarise the exact lengths of the various datasets for both experiments.

The four alternative methods we benchmark against our own Bayesian informative method, rely on four different techniques: the first one is the Bayesian non-informative method that we presented earlier in Section \ref{Pgauss} (recall that it is meant to be an equivalent to the maximum likelihood approach), the second involvedsnon-parametric estimation with kernels \cite[see][]{FanYao}, the third is a double exponential smoothing \cite[see][]{Taylor} and the fourth and last is an ARIMA model. Note that for the second experiment the data available obviously has to be time-reversed in order to apply the last three alternatives methods, since time-dependence plays an important role for them. The ARIMA model is automatically selected \cite[see][]{Hyndman} as the best model with regard to the AIC criterion.

\begin{table}[tbp]
 \begin{center}
 \begin{tabular}{cccc}\hline
 & Estimation \(y^\cA\) & Estimation \(y^\cB\) & Prediction \(y^\cB\) \\ \hline
 Case 1 & 1099 & 125 & 28 \\
 Case 2 & 1099 & 144 & 38 \\ \hline
 \end{tabular}
 \end{center}
 \caption{Sample size (in days) of the datasets for both experiments.}
 \label{tab.datasize}
\end{table}


\begin{table}[tbp]
 \begin{center}
 \begin{tabular}{ccccc}\hline
 & \multicolumn{2}{r}{Case 1} & \multicolumn{2}{r}{Case 2} \\
 & RMSE & MAPE & RMSE & MAPE \\ \hline
informative prior (I) & 770.71 & 3.08 & 2041.70 & 5.71 \\
non-informative prior (NI) & 24440.52 & 100.26 & 7317.56 & 22.03 \\
non-parametric (NP) & 1461.52 & 5.82 & 25091.68 & 77.41 \\
double exp. smoothing (DES) & 1800.12 & 7.41 & 22572.58 & 71.36 \\
linear time series (ARIMA) & 1702.39 & 6.89 & 13839.85 & 43.73 \\ \hline
 \end{tabular}
 \end{center}
 \caption{Overall quality (RMSE in MW, MAPE in \%) of the predictions for both experiments.}
 \label{tab.tableauanne}
\end{table}

It is clear from the results exposed in Table \ref{tab.tableauanne} that the informative prior outperforms all the alternative methods by a large margin in each case. The overall bad performance of the Bayesian non-informative method is not surprising because at least 3 to 4 years of data are usually required to avoid overfitting, for such a parametric model.

\subsection{Role of the hyperparameters}\label{subsec:applihyper}
For this application, the setup is the same as the one described at the start of Section \ref{subsec:applibenchmark}. Our aim here is to point out the role of the hyperparameters introduced within the informative prior and show that besides providing better results than alternative methods as was demonstrated in Section \ref{subsec:applibenchmark}, they also provide a measure of similarity between the short datasets of interest and the dataset used to build the prior. We estimate the model for \(y^\cB\) and \(y^{\cB\prime}\) over the period ranging from 12/01/2009 to 06/30/2010 and predict the next 30 days.

\begin{figure}[p]
 \centering
 \includegraphics[width=.495\textwidth]{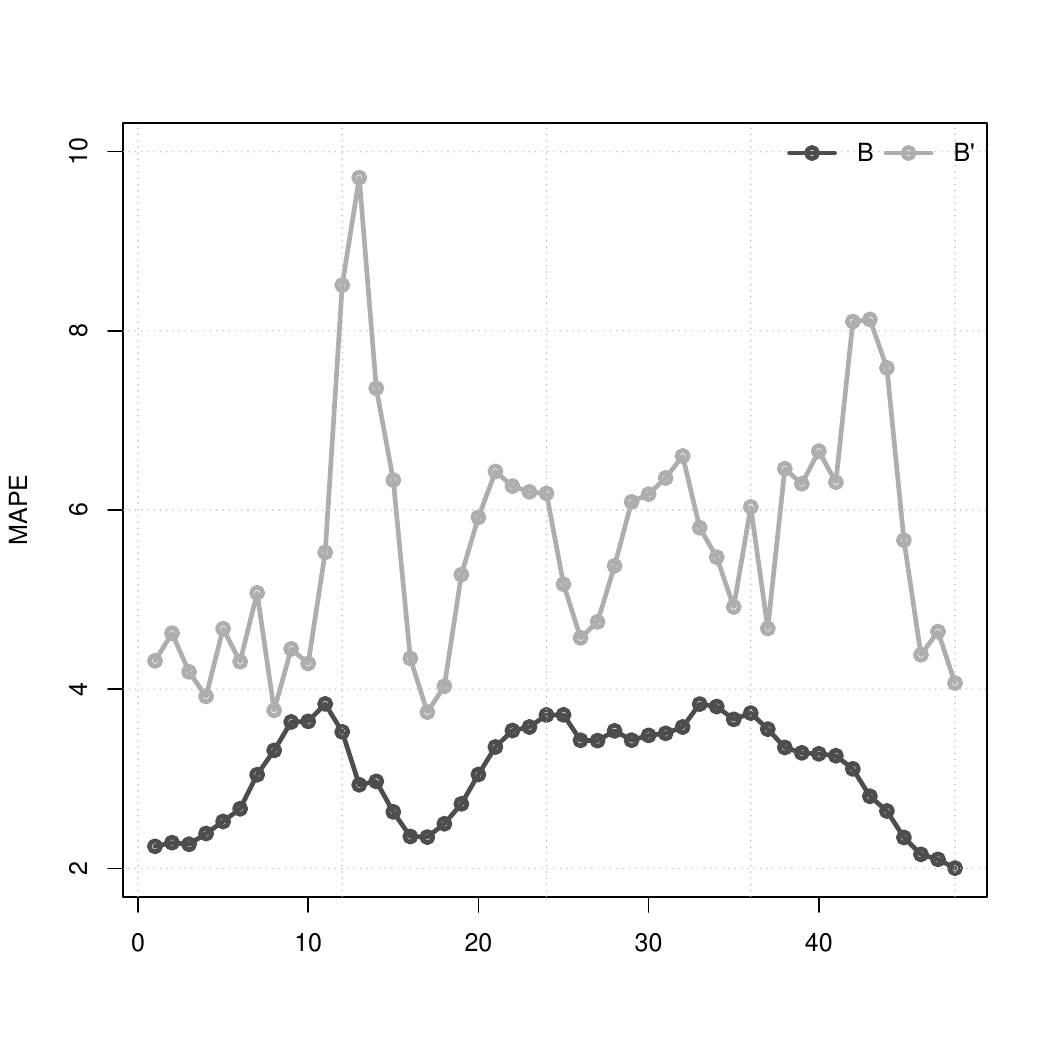}
 \includegraphics[width=.495\textwidth]{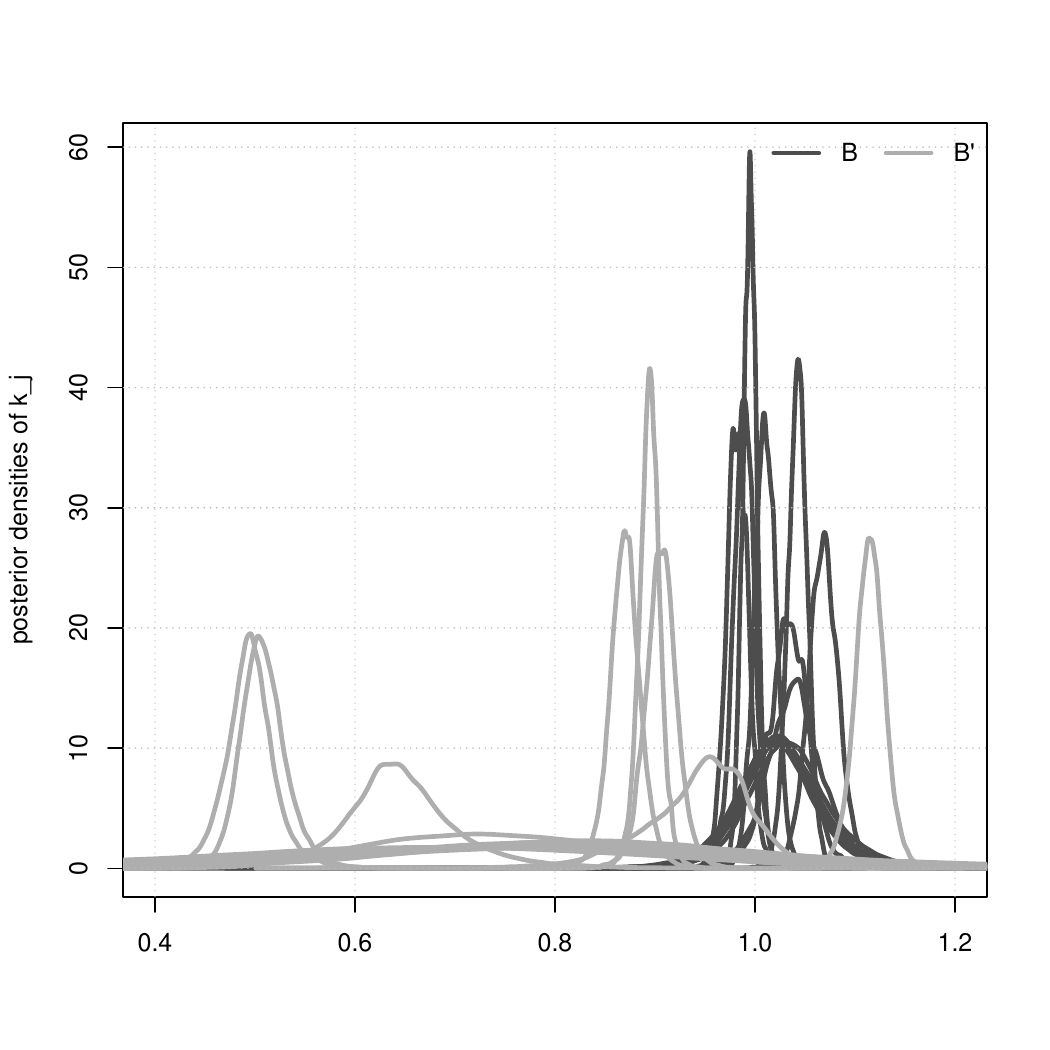}
 \caption{Quality of the predictions (MAPE in \%) averaged for each instant, for populations \(y^\cB\) and \(y^{\cB\prime}\) (left). Posterior densities of the similarity coefficients \(k_j\) for populations \(y^\cB\) and \(y^{\cB\prime}\) at midday (right).}
 \label{fig:Prediction_MAPE}
\end{figure}

\begin{figure}[p]
 \centering
 \includegraphics[width=.495\textwidth]{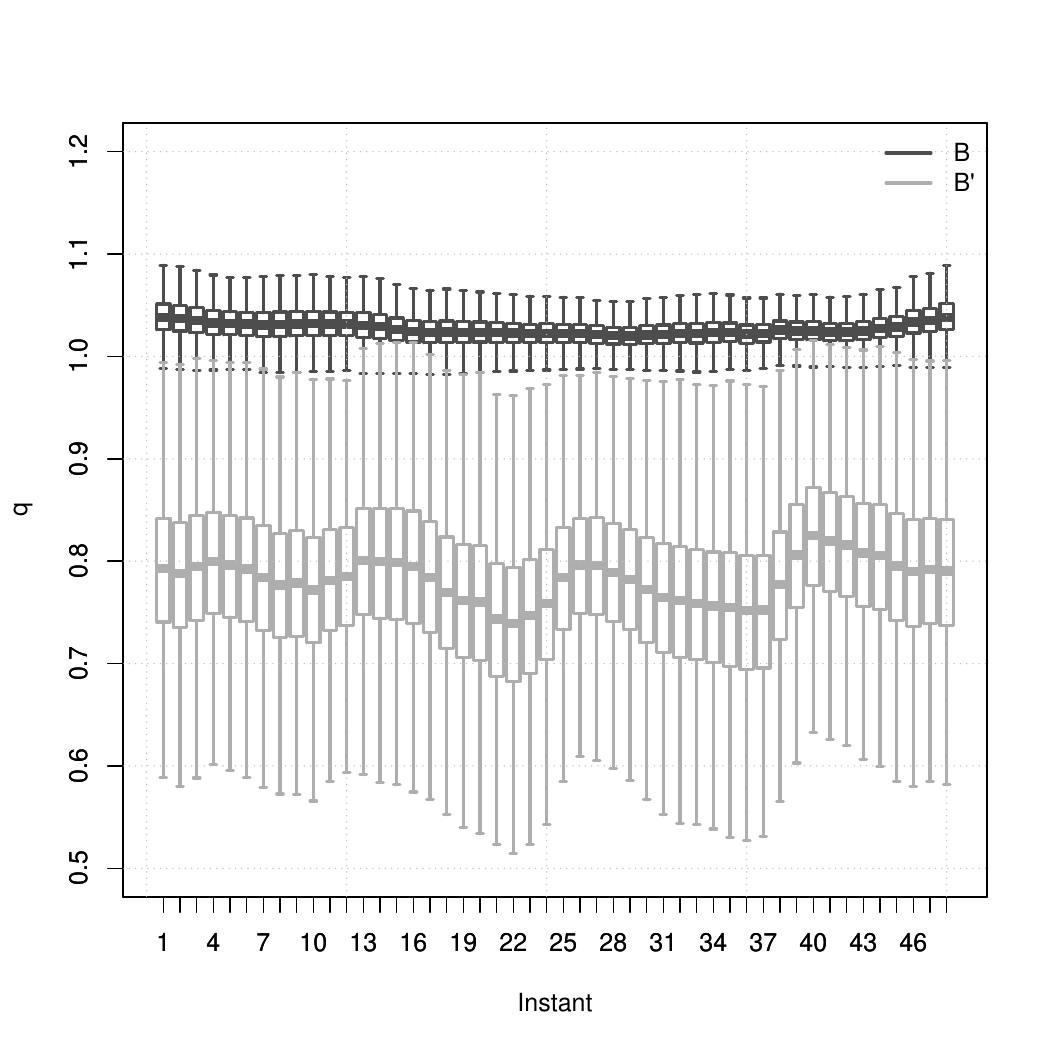}
 \includegraphics[width=.495\textwidth]{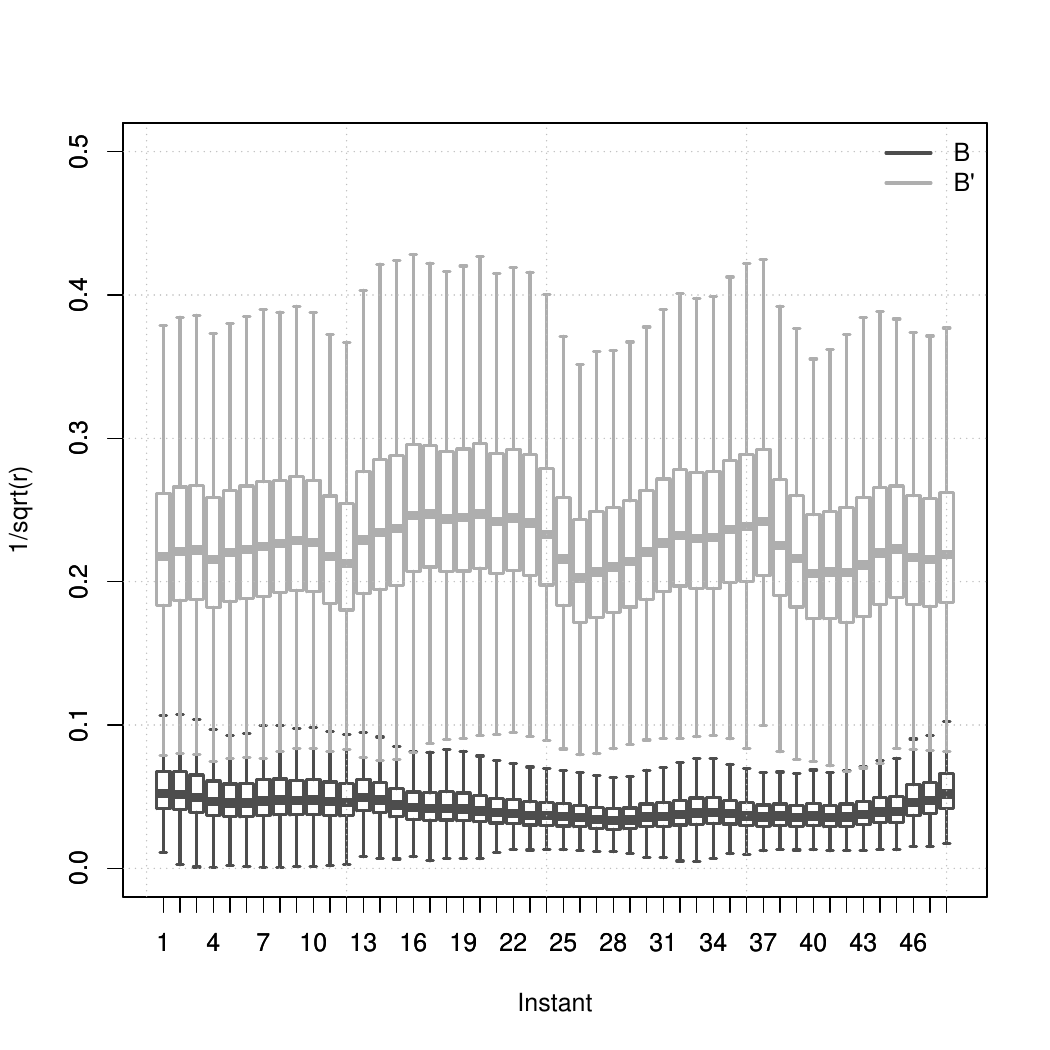}
 \caption{Boxplots of the posterior densities of \(q\) (left) and \(1/\sqrt{r}\) (right) (mean and standard deviation of the similarity coefficients \(k_j\)) for populations \(y^\cB\) and \(y^{\cB\prime}\), throughout the 48 instants of the day.}
 \label{fig:applihyper_qr}
\end{figure}

Figure \ref{fig:Prediction_MAPE} shows the predictive quality of the model for both populations \(y^\cB\) and \(y^{\cB\prime}\) using the informative prior as well as the posterior densities of the similarity coefficients \(k_j\) at midday (we also checked the other 47 instants but the look of them was nearly identical to the one we present). These densities are much more peaked and also centred closer to \(1\) for population \(y^\cB\) than they are for population \(y^{\cB\prime}\). It thus seems to indicate that dataset \(y^\cB\) is more similar to \(y^\cA\) than \(y^{\cB\prime}\) is, confirming our prior knowledge that \(y^\cA\) and \(y^\cB\) cover approximately the same population whereas \(y^{\cB\prime}\) represents around 50\% of \(y^\cA\): this value of 50\% is also visible on Figure \ref{fig:Prediction_MAPE} where we observe two densities centred around \(0.5\), which correspond to the similarity coefficients between the offsets \(\omega_j\) of \(y^\cA\) and these of \(y^{\cB\prime}\).

Figure \ref{fig:applihyper_qr} displays the boxplots for the posterior densities of \(q\) and \(1/\sqrt{r}\) and seem to corroborate the fact that \(y^\cB\) is more similar to \(y^\cA\) than \(y^{\cB\prime}\) is. Recall that \(q\) and \(1/\sqrt{r}\) respectively act as the mean and standard deviation of the similarity coefficients \(k_j\) within our informative hierarchical prior. Indeed the estimated mean of \(q\) appears to be closer to \(1\) while its estimated variance is smaller on \(y^\cB\) than \(y^{\cB\prime}\). The estimated mean and variance of \(1/\sqrt{r}\) are also smaller on \(y^\cB\) than \(y^{\cB\prime}\).

The estimated values of \(q\) and \(r\) provide some information about the similarity between the datasets considered. Notice also that the best predictive performance is obtained when the similarity between the two datasets is strongest.

\subsection{Role of the sample size}\label{subsec:applisamplesize}
For this application, the setup is almost identical to the one described at the start of Section \ref{subsec:applibenchmark} but the temperature used in the model is not the average temperature over France anymore but a transformation of it: it is smoothed using exponential smoothing, which is known to improve the link existing between the two variables temperature and electricity load \cite[see][for more information about this]{Bruhns}. The cooling threshold is fixed at 18$^\circ$C throughout the 48 instants of the day, and this time, the summer holiday break is not removed from the dataset (but the winter holiday break and the bank holidays still are), so that the model could benefit from (and be tested on) the August months in general. Note that for \(y^\cA\) we use the same dataset as we used for our two first applications. 

\begin{figure}[htbp]
 \centering
 \includegraphics{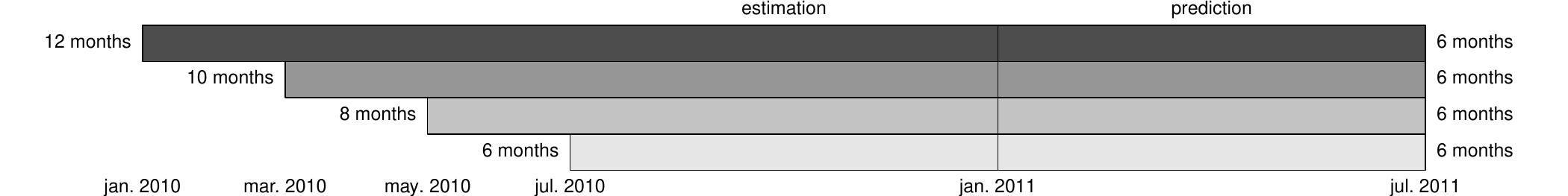}
 \caption{Ranges of the estimation (from 12 to 6 months) and prediction (6 months) time-windows for the 4 scenarios considered.}
 \label{fig:Estimpred_Adelaide}
\end{figure}

We put the focus on the length of the estimation period on \(y^\cB\) while keeping the same prediction window. We successively choose the periods ranging from 01/01/2010, 03/01/2010, 05/01/2010, 07/01/2010 to 12/31/2010, reducing the estimation period on \(y^\cB\) from 12 months to only 6 months, removing 2 months at a time. We then predict the next 6 upcoming months i.e. from 01/01/2011 to 06/30/2011. The diagram in Figure \ref{fig:Estimpred_Adelaide} describes the 4 scenarios considered.

The non-informative prior leads to a better fit than the informative prior as can be seen in Table \ref{tab:qualite_global}. It should not come as a surprise because the non-informative prior was indeed meant to be equivalent to a maximum likelihood approach whose criterion is precisely to minimise the RMSE. As for the quality of the predictions associated with the model for both priors, Table \ref{tab:qualite_global} demonstrates that the informative prior beats the non-informative prior in each of the four proposed configurations. The improvement appears to be minimal when 12 months are used but as months are removed from the estimation window, the predictive quality for the non-informative prior drops very quickly, while the predictive quality for the informative prior remains moderate and stable.

\begin{table}[htbp]
 \begin{center}
 \begin{tabular}{rrrrrrrrr}
 \hline
 & \multicolumn{4}{r}{Estimation} & \multicolumn{4}{r}{Prediction} \\
 & \multicolumn{2}{r}{RMSE} & \multicolumn{2}{r}{MAPE}
 & \multicolumn{2}{r}{RMSE} & \multicolumn{2}{r}{MAPE} \\
 & non-info. & info. & non-info. & info. & non-info. & info. & non-info. & info. \\
 \hline
 12m.& 663.02 & 671.95 & 1.86 & 1.87 & 763.23 & 737.83 & 2.01 & 1.94 \\
 10m.& 606.04 & 623.23 & 1.78 & 1.82 & 1509.09 & 883.07 & 3.18 & 2.21 \\
 8m. & 473.29 & 493.68 & 1.49 & 1.52 & 8891.81 & 1318.28 & 16.72 & 3.26 \\
 6m. & 460.60 & 499.13 & 1.34 & 1.44 & 90356.82 & 1305.27 & 224.40 & 3.62 \\
 \hline
 \end{tabular}
 \end{center}
\caption{Overall quality (RMSE in MW, and MAPE in \%) of the estimation (left) and prediction (right) for the non-informative (non-info.) and informative (info.) priors, depending on the number of months used for the estimation (from 12 months to 6 months).}
 \label{tab:qualite_global}
\end{table}

\begin{figure}[htbp]
 \centering
 \includegraphics[width=.495\textwidth]{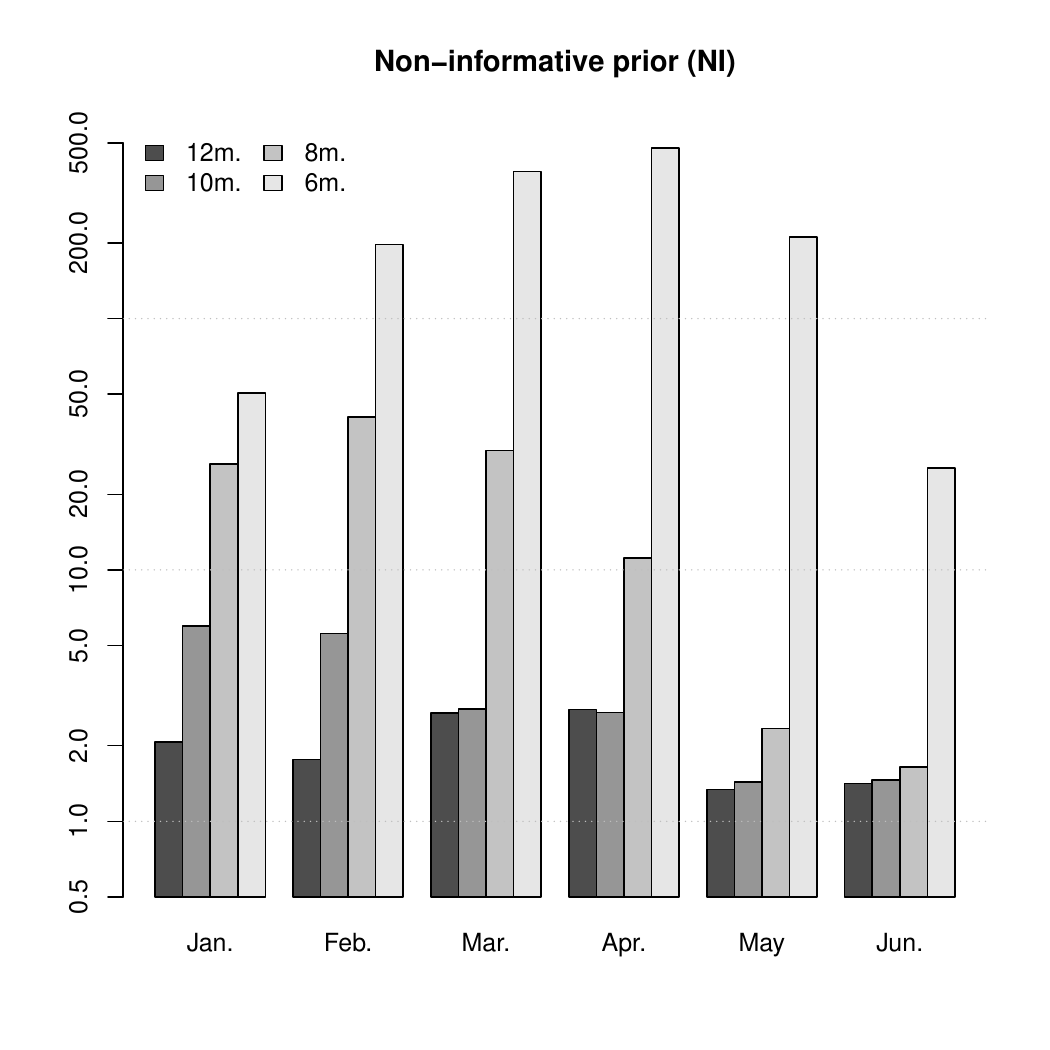}
 \includegraphics[width=.495\textwidth]{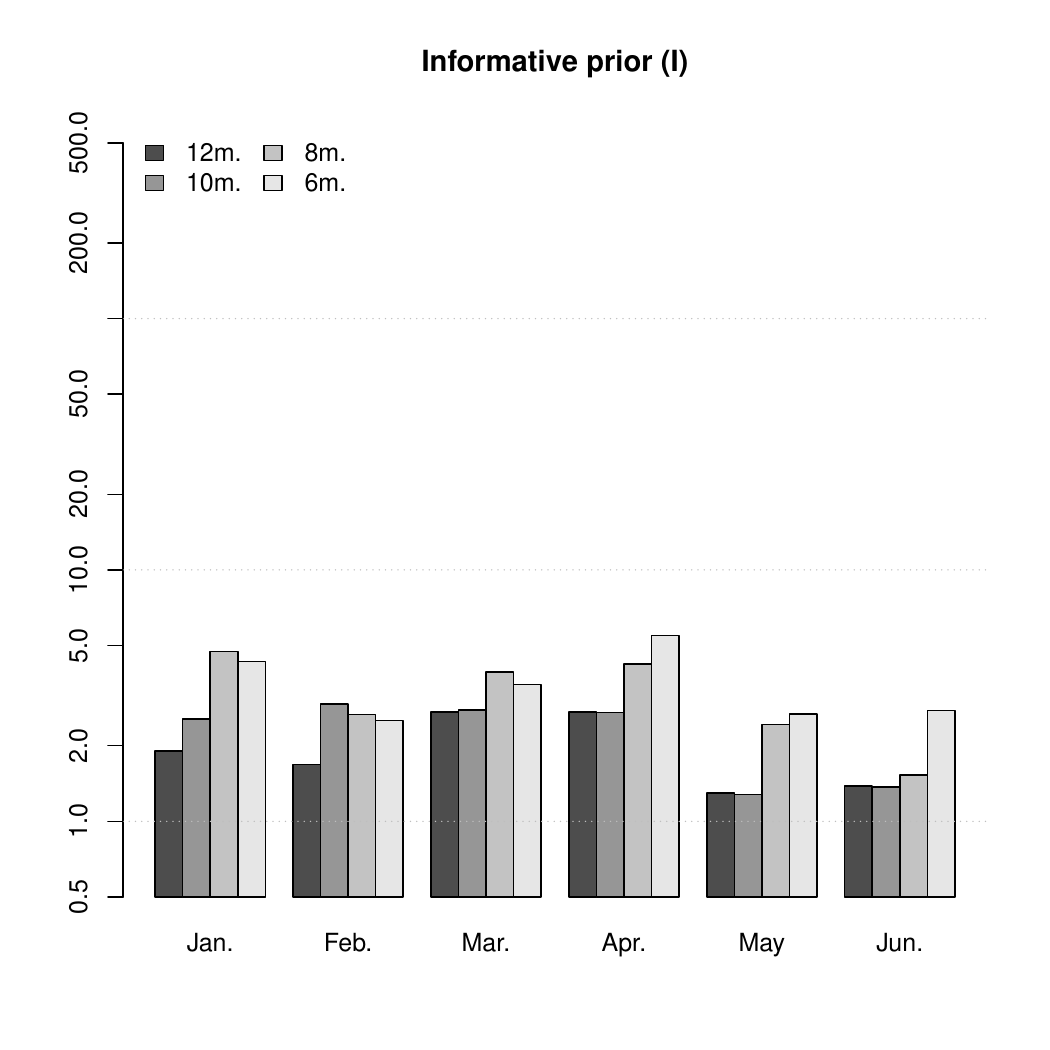}
 \caption{Quality of the predictions (MAPE in \%) averaged for each month (all the instants of the 30 or so days within each month are used to compute these averages), with an estimation period ranging from 12 to 6 months using the non-informative prior (left) and using the informative prior (right). The ordinate axis is in log-scale. Each shade of grey corresponds to a different scenario (as depicted on Figure \ref{fig:Estimpred_Adelaide}).}
 \label{fig:MAPE_Adelaide_mois}
\end{figure}

Figure \ref{fig:MAPE_Adelaide_mois} shows the average error in prediction for each month. It is important to note that the use of the non-informative prior leads to overfitting the model: the results presented in Table \ref{tab:qualite_global} show that as the estimation window goes smaller, the estimation error decreases while the prediction error grows very quickly. A close inspection of the posterior densities of the different parameters of the model reveals that the bias induced by the increasing lack of data is mainly seasonal: this is due to the seasonality coefficients of the model being overfit. Choosing the informative prior over the non-informative prior makes the estimation and prediction of the model more robust with regards to the lack of data.

The informative prior especially improves the quality of the predictions when the lack of data is severe: it provides reasonable forecasts even in the worst scenario considered here, where only 6 months of estimation are used for 6 months of prediction. In this situation, estimation (from 07/01/2010 to 12/31/2010) and prediction (from 01/01/2011 to 06/30/2011) are performed on non-overlapping areas of the calendar: the informative prior makes up for the unavailable data and prevents the model from overfitting on the second half of the calendar while the non-informative prior does not and consequently leads to heavily biased predictions over the first half of the calendar.

%

\section{Appendix}\label{sec:appendix}
\subsection{Notations}Using the notation \(M_{i\bullet}\) for the \(i\)-th row of a matrix \(M\), the non-linear model described in \eqref{eventail} can be re-written in the following condensed way, as a particular case of the regression model \eqref{eq.modeleft}: for \(t = 1,\ldots,N,\)
\begin{align}
y_t &= (A_{t\bullet} \alpha) (B_{t\bullet} \beta + C_{t}) + \gamma (T_t - u)\one_{\interfo{T_t}{+\infty}}(u) + \epsilon_t. \label{eq.modeleelectrique}
\end{align}
The matrices \(A\) of size \(N\times d_A\), \(B\) of size \(N\times d_{\beta}\), \(C\) of size \(N\times 1\), and \(T\) of size \(N\times 1\) are known exogenous variables while the parameters of the model to be estimated are \[\eta = (\theta,\sigma^2) = (\alpha, \beta, \gamma, u, \sigma^2) \in \R^{d_\alpha} \times B_+^{d_\beta}(0, 1) \times \R^* \times \interff{\underline{u}}{\overline{u}} \times \R_+^*,\] where \(B_+^{d_{\beta}}(0, 1) = \{ \beta \in (\R_+)^{d_{\beta}} ;\; \|\beta\|_1 \leq 1 \}\) is the positive quadrant of the \(\|\cdot\|_1\)-unit ball of dimension \(d_{\beta}\).

\subsection{Technical proofs}
\begin{proposition}\label{hierarchicalproperpost}
For \((\beta, u) \in B_+^{d_{\beta}}(0, 1) \times \interff{\underline{u}}{\overline{u}}\) denote \(A_*(\beta, u)\) the matrix whose rows are \[(A_*)_{t\bullet}(\beta, u) = \left[( B_{t\bullet}\beta + C_{t})A_{t\bullet}, (T_{t}-u)\one_{\interfo{T_{t}}{+\infty}}(u)\right], \quad t=1,\ldots,N,\] and suppose \(A_*^\prime(b, u) A_*(b, u)\) has full rank for every \((\beta, u) \in B_+^{d_{\beta}}(0, 1) \times \interff{\underline{u}}{\overline{u}}\). Assume furthermore that \(N>d_\alpha+1\) and that \((y_1,\ldots,y_N)\) are observations coming from the model \eqref{eq.modeleelectrique} and the posterior measure is then a well-defined (proper) probability distribution.
\end{proposition}

\begin{proof}First notice that \( \int {\pi(\eta, k, l, q, r | y)} \ud \sigma^{2} \)
is proportional to
\begin{align*}
\|y - f(\theta)\|_2^{-N} \one_{\interff{0}{1}}(\|\beta\|_1)\one_{\interff{\underline{u}}{\overline{u}}}(u) \pi(\eta|k, l)\pi(k | q, r)\pi(l)\pi(q)\pi(r),
\end{align*}
for almost every \(y\) and that the function \(\theta\mapsto \|y - f(\theta)\|_2^{-N}\) is bounded, for almost every \(y\). The posterior integrability is hence trivial as long as \(\pi(\eta|k, l)\pi(k | q, r)\pi(l)\pi(q)\pi(r)\) itself is a proper distribution which is the case here.
\end{proof}

\begin{proposition}\label{noninfoproperpost}
For \((\beta, u) \in B_+^{d_{\beta}}(0, 1) \times \interff{\underline{u}}{\overline{u}}\) denote \(A_*(\beta, u)\) the matrix whose rows are \[(A_*)_{t\bullet}(\beta, u) = \left[( B_{t\bullet}\beta + C_{t})A_{t\bullet}, (T_{t}-u)\one_{\interfo{T_{t}}{+\infty}}(u)\right], \quad t=1,\ldots,N,\] and suppose \(A_*^\prime(b, u) A_*(b, u)\) has full rank for every \((\beta, u) \in B_+^{d_{\beta}}(0, 1) \times \interff{\underline{u}}{\overline{u}}\). Assume furthermore that \(N>d_\alpha+1\) and that \((y_1,\ldots,y_N)\) are observations coming from the model \eqref{eq.modeleelectrique}, the posterior measure \eqref{noninfopost} is then a well-defined (proper) probability distribution.
\end{proposition}

\begin{proof}Notice first that
\begin{align*}
\int {\pi(\theta, \sigma^2 | y)} \ud \sigma^2 &\propto \|y - f(\theta)\|_2^{-N} \one_{\interff{0}{1}}(\|b\|_1)\one_{\interff{\underline{u}}{\overline{u}}}(u) \quad \text{for almost every } y,
\end{align*}
and observe then that
\begin{align*}
\|y - f(\theta)\|_2^2 &= \sum_{t=1}^N \left[y_{t} - (B_{t\bullet}\beta+C_{t}) A_{t\bullet}\alpha - (T_{t}-u)\one_{\interfo{T_{t}}{+\infty}}(u)\gamma\right]^2.
\end{align*}
Let \((\beta_0, u_0) \in B_+^{d_{\beta}}(0, 1) \times \interff{\underline{u}}{\overline{u}}\) and denote \(\alpha_* = (\alpha, \gamma)\). We write
\begin{align*}
\|y - f((\alpha,\beta_0,\gamma,u_0))\|_2^2 &= \sum_{t=1}^N \left[y_{t} - (B_{t\bullet}\beta_0+C_{t}) A_{t\bullet}\alpha - (T_{t}-u_0)\one_{\interfo{T_{t}}{+\infty}}(u_0)\gamma\right]^2 \\
&= \|y - A_*(\beta_0, u_0) \alpha_*\|_2^2,
\end{align*}
and thus obtain the following equivalence, as \((\beta, u)\rightarrow (\beta_0,u_0)\) and \(\|\alpha_*\|_2\rightarrow +\infty\)
\begin{align}
\|y - f(\theta)\|_2^{-N} \sim \|y - A_*(\beta_0,u_0) \alpha_*\|_2^{-N}.\label{equivb0u0}
\end{align}
The triangular inequality applied to the right hand side of \eqref{equivb0u0} gives
\begin{align}
\|y - A_*(\beta_0,u_0) \alpha_*\|_2^{-N} &\leq \big| \|y\|_2 - \|A_*(\beta_0,u_0) \alpha_*\|_2 \big|^{-N}.\label{triangineqb0u0}
\end{align}
Since \(A_*^\prime(\beta_0,u_0) A_*(\beta_0,u_0)\) has full rank, by straightforward algebra we get
\begin{align*}
\lambda \|\alpha_*\|_2^2 &\leq \|A_* (\beta_0,u_0)\alpha_*\|_2^2,
\end{align*}
where \(\lambda\) is the smallest eigenvalue \((A_*(\beta_0,u_0))^\prime A_*(\beta_0,u_0)\) and is strictly positive. We can hence find an equivalent of the right hand side of \eqref{triangineqb0u0} as \(\|\alpha_*\|_2\rightarrow +\infty\), which is
\begin{align}
\big| \|y\|_2 - \|A_*(\beta_0,u_0) \alpha_*\|_2 \big|^{-N} \sim \lambda^{-N/2} \|\alpha_*\|_2^{-N}.\label{equivlambda}
\end{align}
Combining \eqref{equivb0u0}, \eqref{triangineqb0u0} and \eqref{equivlambda} together, we see that the integrability of the left hand side of \eqref{equivb0u0} as \((\beta, u)\rightarrow (\beta_0,u_0)\) and \(\|\alpha_*\|_2\rightarrow +\infty\) is directly implied by that of \(\|\alpha_*\|_2^{-N}\). The latter is immediate for \(N > d_\alpha+1\), as can be seen via a Cartesian to hyperspherical re-parametrisation.

The previous paragraph thus ensures the integrability of \(\|y - f(\theta)\|_2^{-N}\) over sets of the form \[\{(\beta,u)\in V((\beta_0, u_0)), \|\alpha_*\|_2 \in \interoo{M(\beta_0,u_0)}{+\infty}\}, \quad \forall (\beta_0, u_0) \in B_+^{d_{\beta}}(0, 1) \times \interff{\underline{u}}{\overline{u}}\] where the subset \(V((b_0, u_0))\) is an open neighbourhood of \((\beta_0, u_0)\) and \(M(\beta_0, u_0)\) is a real number depending on \((\beta_0, u_0)\). By compacity of \(B_+^{d_{\beta}}(0, 1) \times \interff{\underline{u}}{\overline{u}}\) there exists a finite union of such \(V((\beta_i, u_i))\) that covers \(B_+^{d_{\beta}}(0, 1) \times \interff{\underline{u}}{\overline{u}}\). Denoting \(M\) the maximum of \(M(\beta_i, u_i)\) over the corresponding finite subset of \((\beta_i, u_i)\), we finally obtain the integrability of \(\|y - f(\theta)\|_2^{-N}\) over \(\{(\beta,u)\in B_+^{d_{\beta}}(0, 1), \|\alpha_*\| \in \interoo{M}{+\infty}\}\).

The integrability of \(\|y - f(\theta)\|_2^{-N}\) over \(\{(\beta,u)\in B_+^{d_{\beta}}(0, 1), \|\alpha_*\| \in \interff{0}{M}\}\) is trivial, recalling that \(\theta \mapsto \|y - f(\theta)\|_2\) is continuous and does not vanish over this compact for almost every \(y\), meaning its inverse shares these same properties.
\end{proof}

\begin{remark}
The condition ``\(A_*^\prime A_*\) has full rank'' mentioned above is typically verified in our applications for the regressors used in our model. To see this, call ``vector of heating degrees'' the vector whose coordinates are \((T_{t}-u)\one_{\interfo{T_{t}}{+\infty}}(u)\), then not verifying the aforementioned condition is equivalent to saying that there exists an index \(i\) and a threshold \(u\) such that the family of vectors formed by the regressors \(A\) and the vector of heating degrees is linearly dependent over the subset \(\Psi_i\) of the calendar''.
\end{remark}

\subsection{MCMC algorithms for the estimation of the posterior density}
The two MCMC algorithms presented below were developed because direct simulations from the posterior distribution were not possible. The justifications are given after the algorithms themselves. Notice that the full conditional distributions of all the parameters but the threshold \(u\) appear to be common distributions in both cases, due to the presence of multiple semi-conjugacy situations. We used a Metropolis-within-Gibbs algorithm \cite[see][page 230, for a description]{RobertCasella2} based on Gibbs sampling steps for every parameter but \(u\) for which we used a Metropolis-Hasting step based on a Gaussian random walk proposal.

\subsubsection{MCMC algorithm for the estimation of the posterior distribution, using the informative prior}
In the lines below we give the different steps of the MCMC algorithm we used to (approximately) simulate \((\eta_1, \ldots, \eta_M)\) according to the posterior distribution \({\pi(\eta|y)}\) corresponding to the informative prior we presented earlier. Denoting \(\loi{N}(\mu, \Sigma, S)\) the \(S\)-truncated Gaussian distribution with mean \(\mu\) and covariance \(\Sigma\), the algorithm goes as follow:

{\ttfamily
\begin{enumerate}[Step 1.]
 \item Initialise \(\eta_1\) such that \({\pi(\eta_1|y)} \neq 0\)
 \item For \(t=1,\ldots,M-1\), repeat
 \item[(i).] Simulate \(\sigma^2_{t+1} \text{ cond. to } (\alpha_{t}, \beta_{t}, \gamma_{t}, u_{t}, k_{t}, l_{t}, q_{t}, r_{t}, y) \)
 \[\sigma^2_{t+1} \sim \loi{IG}\left(\frac{N}{2}, \frac{1}{2}\|y - f(\theta)\|_2^2\right)\]
 \item[(ii).] Simulate \(r_{t+1} \text{ cond. to } (\alpha_{t}, \beta_{t}, \gamma_{t}, u_{t}, \sigma^2_{t+1}, k_{t}, l_{t}, q_{t}, y) \)
 \[ r_{t+1} \sim \loi{G}\left(a_r+\frac{d}{2}, b_r+\frac{1}{2}\sum_{i=1}^d (k_i-q)^2\right)\]
 \item[(iii).] Simulate \(q_{t+1} \text{ cond. to } (\alpha_{t}, \beta_{t}, \gamma_{t}, u_{t}, \sigma^2_{t+1}, k_{t}, l_{t}, r_{t+1}, y) \)
 \[ q_{t+1} \sim \loi{N}\left([\sigma_q^{-2}+rd]^{-1}(\sigma_q^{-2} + r \sum_{i=1}^d k_i), [\sigma_q^{-2}+rd]^{-1}\right) \]
 \item[(iv).] Simulate \(l_{t+1} \text{ cond. to } (\alpha_{t}, \beta_{t}, \gamma_{t}, u_{t}, \sigma^2_{t+1}, k_{t}, q_{t+1}, r_{t+1}, y) \)
 \[ l_{t+1} \sim \loi{G}\left(a_l+\frac{d}{2}, b_l+\frac{1}{2}(\eta_{t}-K\mu^{\cA}_{t})^\prime (\Sigma^{\cA})^{-1}(\eta_{t}-K\mu^{\cA}_{t})\right)\]
 \item[(v).] Simulate \(k_{t+1} \text{ cond. to } (\alpha_{t}, \beta_{t}, \gamma_{t}, u_{t}, \sigma^2_{t+1}, l_{t+1}, q_{t+1}, r_{t+1}, y) \)
 \[k_{t+1} \sim \loi{N}\left(\mu_{t+1}^k, \Sigma_{t+1}^k\right)\]
 \item[(vi).] Simulate \(\gamma_{t+1} \text{ cond. to } (\alpha_{t}, \beta_{t}, u_{t}, \sigma^2_{t+1}, k_{t+1}, l_{t+1}, q_{t+1}, r_{t+1}, y) \)
 \[\gamma_{t+1} \sim \loi{N}\left(\mu_{t+1}^g, \Sigma_{t+1}^g\right)\]
 \item[(vii).] Simulate \(\beta_{t+1} \text{ cond. to } (\alpha_{t}, \gamma_{t+1}, u_{t}, \sigma^2_{t+1}, k_{t+1}, l_{t+1}, q_{t+1}, r_{t+1}, y) \)
 \[\beta_{t+1} \sim \loi{N}\left(\mu_{t+1}^b, \Sigma_{t+1}^b, B_+^{d_{\beta}}(0, 1)\right)\]
 \item[(viii).] Simulate \(\alpha_{t+1} \text{ cond. to } (\beta_{t+1}, \gamma_{t+1}, u_{t}, \sigma^2_{t+1}, k_{t+1}, l_{t+1}, q_{t+1}, r_{t+1}, y) \)
 \[\alpha_{t+1} \sim \loi{N}\left(\mu_{t+1}^a, \Sigma_{t+1}^a\right)\]
 \item[(ix).] Simulate \(\delta_{t} \sim \loi{N}(0, \Sigma_{\textrm{MH}}),\quad \upsilon_{t} \sim \loi{U}\interff{0}{1}\) and define \(\widetilde{u}_{t} = u_{t} + \delta_{t}\)
 \begin{itemize}
 \item define \(u_{t+1} = \widetilde{u}_{t}\) if \[\upsilon_{t} < \frac{\pi(\widetilde{u}_{t}|\alpha_{t+1}, \beta_{t+1}, \gamma_{t+1}, \sigma^2_{t+1}, k_{t+1}, l_{t+1}, q_{t+1}, r_{t+1}, y)}{\pi(u_{t}|\alpha_{t+1}, \beta_{t+1}, \gamma_{t+1}, \sigma^2_{t+1}, k_{t+1}, l_{t+1}, q_{t+1}, r_{t+1}, y)}\]
 \item or \(u_{t+1} = u_{t}\) otherwise
 \end{itemize}
\end{enumerate}}
\noindent where the covariance matrix \(\Sigma_{\text{MH}}\) used in this last Metropolis-Hastings step is first estimated over a burn-in phase (the iterations coming from this phase are discarded), and then fixed to its estimated value ``asymptotically optimally rescaled'' for the final run by a factor \(\left(\frac{2.38}{d}\right)^2\) \cite[as recommended for Gaussian proposals in Section 2 of][]{RobertsOptimal}.

The justifications for each full conditional distribution used in the Gibbs sampling steps, including the explicit expressions of \(\mu_{t+1}^\alpha, \Sigma_{t+1}^\alpha, \mu_{t+1}^\beta, \Sigma_{t+1}^\beta, \mu_{t+1}^\gamma, \Sigma_{t+1}^\gamma, \mu_{t+1}^k\) and \(\Sigma_{t+1}^k\), are to be found in \cite{LaunayPhD}.

\subsubsection{MCMC algorithm for the estimation of the posterior distribution, using the non-informative prior}
In the lines below, we give the different steps of the MCMC algorithm we used to (approximately) simulate \((\eta_1, \ldots, \eta_M)\) according to the posterior distribution \({\pi(\eta|y)}\) corresponding to the non-informative prior we presented earlier. Denoting \(\loi{N}(\mu, \Sigma, S)\) the \(S\)-truncated Gaussian distribution with mean \(\mu\) and covariance \(\Sigma\), the algorithm goes as follows:

{\ttfamily
\begin{enumerate}[Step 1.]
 \item Initialise \(\eta_1\) such that \({\pi(\eta_1|y)} \neq 0\)
 \item For \(t=1,\ldots,M-1\), repeat
 \item[(i).] Simulate \(\sigma^2_{t+1} \text{ cond. to } (\alpha_{t}, \beta_{t}, \gamma_{t}, u_{t}, y) \) i.e. \[\sigma^2_{t+1} \sim \loi{IG}\left(\frac{N}{2}, \frac{1}{2}\|y - f(\eta)\|_2^2\right)\]
 \item[(ii).] Simulate \(\gamma_{t+1} \text{ cond. to } (\alpha_{t}, \beta_{t}, u_{t}, \sigma^2_{t+1}, y) \) i.e. \[\gamma_{t+1} \sim \loi{N}\left(\mu_{t+1}^\gamma, \Sigma_{t+1}^\gamma\right)\]
 \item[(iii).] Simulate \(b_{t+1} \text{ cond. to } (\alpha_{t}, \gamma_{t+1}, u_{t}, \sigma^2_{t+1}, y) \) i.e. \[\beta_{t+1} \sim \loi{N}\left(\mu_{t+1}^\beta, \Sigma_{t+1}^\beta, B_+^{d_{\beta}}(0, 1)\right)\]
 \item[(iv).] Simulate \(a_{t+1} \text{ cond. to } (\beta_{t+1}, \gamma_{t+1}, u_{t}, \sigma^2_{t+1}, y) \) i.e. \[\alpha_{t+1} \sim \loi{N}\left(\mu_{t+1}^\alpha, \Sigma_{t+1}^\alpha\right)\]
 \item[(v).] Simulate \(\delta_{t} \sim \loi{N}(0, \Sigma_{\textrm{MH}})\), simulate \(\upsilon_{t} \sim \loi{U}\interff{0}{1}\) and define \(\widetilde{u}_{t} = u_{t} + \delta_{t}\)
 \begin{itemize}
 \item define \(u_{t+1} = \widetilde{u}_{t}\) if \[\upsilon_{t} < \frac{\pi(\widetilde{u}_{t}|\alpha_{t+1}, \beta_{t+1}, \gamma_{t+1}, \sigma^2_{t+1}, y)}{\pi(u_{t}|\alpha_{t+1}, \beta_{t+1}, \gamma_{t+1}, \sigma^2_{t+1}, y)}\]
 \item or \(u_{t+1} = u_{t}\) otherwise
 \end{itemize}
\end{enumerate}}
\noindent where the covariance matrix \(\Sigma_{\text{MH}}\) used in this last Metropolis-Hastings step is first estimated over a burn-in phase (the iterations coming from this phase are discarded), and then fixed to its estimated value ``asymptotically optimally rescaled'' for the final run by a factor \(\left(\frac{2.38}{d}\right)^2\) \cite[as recommended for Gaussian proposals in Section 2 of][]{RobertsOptimal}. 

The justifications for each full conditional distribution used in the Gibbs sampling steps, including the explicit expressions of \(\mu_{t+1}^\alpha, \Sigma_{t+1}^\alpha, \mu_{t+1}^\beta, \Sigma_{t+1}^\beta, \mu_{t+1}^\gamma, \text{ and } \Sigma_{t+1}^\gamma\), are to be found in \cite{LaunayPhD}.

\section*{Acknowledgments}
The authors would like to thank Ad\'{e}la\"{i}de Priou for collecting a part of the data as well as the corresponding results, and Virginie Dordonnat for the insightful discussions.

\bibliographystyle{apalike}

\begin{thebibliography}{}

\end{thebibliography}


\begin{thebibliography}{}

\bibitem[Abramovitz and Stegun, 1965]{abramovitzstegun65}
Abramovitz, M. and Stegun, I. (1965).
\newblock {\em Handbook of {M}athematical {F}unctions with {F}ormulas,
  {G}raphs, and {M}athematical {T}ables}.
\newblock New York, Dover Publications.

\bibitem[Al-Zayer and Al-Ibrahim, 1996]{Alzayer}
Al-Zayer, J. and Al-Ibrahim, A. (1996).
\newblock {Modelling the Impact of Temperature on Electricity Consumption in
  the Eastern Province of Saudi Arabia}.
\newblock {\em Journal of Forecasting}, 15:97--106.

\bibitem[Albert, 2009]{albert2009Bayesian}
Albert, J. (2009).
\newblock {\em Bayesian computation with R}.
\newblock Springer, Dordrecht.

\bibitem[Bouveyron and Jacques, 2013]{bouveyron2010adaptive}
Bouveyron, C. and Jacques, J. (2013).
\newblock Adaptive mixtures of regressions: Improving predictive inference when
  population has changed.
\newblock {\em Communications in Statistics - Simulation and Computation, to
  appear}.

\bibitem[Bruhns et~al., 2005]{Bruhns}
Bruhns, A., Deurveilher, G., and Roy, J. (2005).
\newblock {A Non-Linear Regression Model for Mid-Term Load Forecasting and
  Improvements in Seasonnality}.
\newblock {\em Proceedings of the 15th Power Systems Computation Conference
  2005, Liege Belgium}.

\bibitem[Bunn and Farmer, 1985]{BunnFarmer}
Bunn, D. and Farmer, E. (1985).
\newblock {\em {Comparative Models For Electrical Load Forecasting}}.
\newblock John Wiley, New York.

\bibitem[Congdon, 2010]{congdon2010}
Congdon, P. (2010).
\newblock {\em Applied Bayesian Hierarchical Methods}.
\newblock Applied Bayesian Hierarchical Methods, Chapman \& Hall, CRC.

\bibitem[Cottet and Smith, 2003]{Cottet}
Cottet, R. and Smith, M. (2003).
\newblock {Bayesian modeling and forecasting of intraday electricity load}.
\newblock {\em Journal of the American Statistical Association},
  98(464):839--849.

\bibitem[Cugliari, 2011]{Cugliari}
Cugliari, J. (2011).
\newblock {\em {Pr\'evision non param\'etrique de processus \`a valeurs
  fonctionnelles, application \`a la consommation d'\'electricit\'e}}.
\newblock PhD thesis, Univ\'ersit\'e Paris Sud XI.

\bibitem[Dordonnat et~al., 2008]{Dordonnat}
Dordonnat, V., Koopman, S., Ooms, M., Dessertaine, A., and Collet, J. (2008).
\newblock {An Hourly Periodic State Space Model for Modelling French National
  Electricity Load}.
\newblock {\em International Journal of Forecasting}, 24(4):566--587.

\bibitem[Efron, 2010]{Efron-empirical-Bayes}
Efron, B. (2010).
\newblock {\em {Large-Scale Inference: Empirical Bayes Methods for Estimation,
  Testing, and Prediction (Institute of Mathematical Statistics Monographs)}}.
\newblock Cambridge University Press.

\bibitem[Engle et~al., 1986]{EngleGrangerRice}
Engle, R., Granger, C., Rice, J., and Weiss, A. (1986).
\newblock Semiparametric estimates of the relation between weather and
  electricity.
\newblock {\em Journal of the American Statistical Association}, 81:310--320.

\bibitem[Fan and Yao, 2005]{FanYao}
Fan, J. and Yao, Q. (2005).
\newblock {\em Non linear Time Series: Nonparametric and Parametric Methods}.
\newblock Springer.

\bibitem[Gelman et~al., 2013]{Gelmanetal}
Gelman, A., Carlin, J., Stern, H., Dunson, D., Vehtari, A., and Rubin, D.
  (2013).
\newblock {\em Bayesian Data Analysis}.
\newblock Chapman \& Hall CRC Texts in Statistical Science.

\bibitem[Gelman and Hill, 2007]{GelmanHill}
Gelman, A. and Hill, J. (2007).
\newblock {\em Data Analysis Using Regression and Multilevel/Hierarchical
  Models}.
\newblock Cambridge University Press.

\bibitem[Ghosh et~al., 2006]{Ghosh}
Ghosh, J.~K., Delampady, M., and Samanta, T. (2006).
\newblock {\em An Introduction to Bayesian Analysis, Theory and Methods}.
\newblock Springer.

\bibitem[Harrison and Stevens, 1976]{HarrisonStevens}
Harrison, P. and Stevens, C. (1976).
\newblock Bayesian forecasting.
\newblock {\em Journal of the Royal Statistical Society}, 38(3):205--247.

\bibitem[Hyndman and Khandakar, 2008]{Hyndman}
Hyndman, R.~J. and Khandakar, Y. (2008).
\newblock Automatic time series forecasting: the forecast package for {R}.
\newblock {\em Journal of Statistical Software}, 88(3).

\bibitem[Launay, 2012]{LaunayPhD}
Launay, T. (2012).
\newblock {\em Bayesian methods for electricity load forecasting}.
\newblock PhD thesis, Universit\'e de Nantes.

\bibitem[Launay et~al., 2012]{Launay2}
Launay, T., Philippe, A., and Lamarche, S. (2012).
\newblock {Consistency of the posterior distribution and MLE for piecewise
  linear regression}.
\newblock {\em Electron. J. Statist.}, 6:1307--1357.

\bibitem[Marin and Robert, 2014]{MarinRobert}
Marin, J.-M. and Robert, C. (2014).
\newblock {\em Bayesian Essentials with R}.
\newblock Texts in Statistics. Springer, 2nd edition edition.

\bibitem[Menage et~al., 1988]{Menage}
Menage, J.~P., Panciatici, P., and Boury, F. (1988).
\newblock Nouvelle modelisation de l'influence des conditions climatiques sur
  la consommation d'energie electrique.
\newblock Technical report, EDF R\&D.

\bibitem[Minka, 1999]{Minka}
Minka, T.~P. (1999).
\newblock Bayesian linear regression.
\newblock Technical report, 3594 Security Ticket Control.

\bibitem[Ramanathan et~al., 1997]{RamanathanEngle}
Ramanathan, R., Engle, R., Granger, C., Vahid-Araghi, F., and Brace, C. (1997).
\newblock Short-run forecasts of electricity loads and peaks.
\newblock {\em International Journal of Forecasting}, 13:161--174.

\bibitem[Robert and Casella, 2009]{RobertCasella2}
Robert, C.~P. and Casella, G. (2009).
\newblock {\em Introducing Monte Carlo Methods with R}.
\newblock Springer Verlag, 1st edition.

\bibitem[Roberts and Rosenthal, 2001]{RobertsOptimal}
Roberts, G.~O. and Rosenthal, J.~S. (2001).
\newblock Optimal scaling for various {Metropolis-Hastings} algorithms.
\newblock {\em Statistical Science}, 16(4):351--367.

\bibitem[Seber and Wild, 2003]{SeberWild}
Seber, G. A.~F. and Wild, C.~J. (2003).
\newblock {\em Nonlinear Regression (Wiley Series in Probability and
  Statistics)}.
\newblock Wiley-Interscience.

\bibitem[Smith, 2000]{Smith}
Smith, M. (2000).
\newblock Modeling and short-term forecasting of new south wales electricity
  system load.
\newblock {\em Journal of Business \& Economic Statistics}, 18:465--478.

\bibitem[Soares and Medeiros, 2008]{SoaresMedeiros}
Soares, L. and Medeiros, M. (2008).
\newblock Modeling and forecasting short-term electricity load: a comparison of
  methods with an application to brazilian data.
\newblock {\em International Journal of Forecasting}, 24:630--644.

\bibitem[Taylor, 2003]{Taylor}
Taylor, J.~W. (2003).
\newblock Short-term electricity demand forecasting using double seasonal
  exponential smoothing.
\newblock {\em Journal of the Operational Research Society}, 54(8):799--805.

\bibitem[Taylor and Buizza, 2003]{TaylorBuizza}
Taylor, J.~W. and Buizza, R. (2003).
\newblock Using weather ensemble predictions in electricity demand forecasting.
\newblock {\em International Journal of Forecasting}, 19(1):57--70.

\bibitem[Yang and Berger, 1998]{Yangberger98}
Yang, R. and Berger, J. (1998).
\newblock A catalog of noninformative priors.
\newblock Technical report, Institute of Statistics and Decision Sciences, Duke
  University.

\end{thebibliography}

\end{document}